\documentclass[11pt]{article}

\pdfoutput=1
\usepackage{amsmath}
\usepackage{amsthm}
\usepackage{amsfonts}          
\usepackage{amssymb}           
\usepackage{color}
\usepackage{mathtools}
\usepackage{cite}
\usepackage{multirow,setspace,microtype}
\usepackage{bm}
\usepackage{booktabs}
\usepackage{tikz}

\definecolor{darkblue}{rgb}{0.0,0.0,0.3} 		

\usepackage[colorlinks]{hyperref}
\hypersetup{pdftitle={Multiple Line Bundle Hidden Sectors for the B-L MSSM},
	pdfauthor={Anthony Ashmore, Sebastian Dumitru and Burt Ovrut},
	linkcolor=darkblue, urlcolor=darkblue, citecolor=darkblue, filecolor=black, linktoc=page}

\PassOptionsToPackage{linecolor=blue,backgroundcolor=blue!25,bordercolor=blue,textsize=scriptsize}{todonotes}
\usepackage{todonotes}

\numberwithin{equation}{section}

\usepackage{makecell}

\usepackage[bf]{caption}
\let\originalleft\left
\let\originalright\right
\renewcommand{\left}{\mathopen{}\mathclose\bgroup\originalleft}
\renewcommand{\right}{\aftergroup\egroup\originalright}

\makeatletter
\g@addto@macro\bfseries{\boldmath}
\makeatother

\newlength{\xtrawidth}
\newlength{\xtraheight}
\renewcommand{\(}{\left(}
\renewcommand{\)}{\right)}
\newcommand{\Z}{\ensuremath{\mathbb{Z}}}

\newcommand{\Rhat}{\ensuremath{{\widehat R}}}

\global\long\def\rep#1{\boldsymbol{\underline{#1}}}%
\global\long\def\repb#1{\overline{\boldsymbol{\underline{#1}}}}%
\global\long\def\ii{\text{i}}%
\global\long\def\ee{\text{e}}%
\global\long\def\SO#1{{SO}(#1)}%
\global\long\def\Uni#1{{U}(#1)}%
\global\long\def\SU#1{{SU}(#1)}%
\global\long\def\Ex#1{{E}_{#1}}%
\global\long\def\ex#1{\mathfrak{e}_{#1}}%
\global\long\def\kernel{\operatorname{ker}}%
\global\long\def\eqspace{\mathrel{\phantom{{=}}{}}}%
\global\long\def\op#1{\operatorname{#1}}%

\DeclareMathOperator{\tr}{tr}

\DeclareMathOperator{\rank}{rk}

\global\long\def\id{\operatorname{id}}%

\global\long\def\tr{\operatorname{tr}}%
\global\long\def\id{\operatorname{id}}%
\global\long\def\kernel{\operatorname{ker}}%
\global\long\def\eqspace{\mathrel{\phantom{{=}}{}}}%
\global\long\def\op#1{\operatorname{#1}}%
\global\long\def\bR{\mathbb{R}}%

\usepackage{subcaption}
\usepackage{float}
\usepackage{caption}

\begin{document}

\begin{titlepage}
\title{\LARGE \bf{Hidden Sectors from Multiple Line Bundles{\\[.15cm]  }for the $B-L$ MSSM}\\[.3cm]}
                       
\author{{
   Anthony Ashmore,$^{a,b}$
   Sebastian Dumitru$^{c}$
   and Burt A.\,Ovrut$^{c}$}\\[0.8cm]
{${}^a$\it Enrico Fermi Institute \& Kadanoff Center for Theoretical Physics} \\[.1cm]
{\it University of Chicago, Chicago, IL 60637, USA}    \\[0.4cm] 
{${}^b$\it Sorbonne Universit\'e, LPTHE,} \\[.1cm]
{\it  F-75005 Paris, France}\\[0.4cm] 
   {${}^c$\it Department of Physics, University of Pennsylvania} \\[.1cm]
   {\it Philadelphia, PA 19104, USA}
 }      

\date{}

\maketitle

\begin{abstract}
\noindent
We give a formalism for constructing hidden sector bundles as extensions of sums of line bundles in heterotic $M$-theory. Although this construction is generic, we present it within the context of the specific Schoen threefold that leads to the physically realistic $B-L$ MSSM model. We discuss the embedding of the line bundles, the existence of the extension bundle, and a number of necessary conditions for the resulting bundle to be slope-stable and thus $N=1$ supersymmetric. An explicit example is presented, where two line bundles are embedded into the $SU(3)$ factor of the $E_{6} \times SU(3)$ maximal subgroup of the hidden sector $E_{8}$ gauge group, and then enhanced to a non-Abelian $SU(3)$ bundle by extension. For this example, there are in fact six inequivalent extension branches, significantly generalizing that space of solutions compared with hidden sectors constructed from a single line bundle.

\noindent

\let\thefootnote\relax\footnotetext{\noindent ashmore@uchicago.edu, sdumitru@sas.upenn.edu, ovrut@elcapitan.hep.upenn.edu}

\end{abstract}

\thispagestyle{empty}
\end{titlepage}

\tableofcontents

\section{Introduction}

Ho\v{r}ava and Witten were the first to put eleven-dimensional $M$-theory on a $S^{1}/\mathbb{Z}_{2}$ orbifold~\cite{Horava:1995qa,Horava:1996ma}. Consistency of the theory requires a ten-dimensional orbifold plane at each end of the $S^{1}/\mathbb{Z}_{2}$ interval, with each plane carrying an $E_{8}$ gauge connection. The degrees of freedom arising from each orbifold plane are usually referred to as the ``observable'' and ``hidden'' sectors respectively. One can compactify Ho\v{r}ava--Witten theory on a Calabi--Yau threefold to give five-dimensional ``heterotic $M$-theory''. Integrating out the modes on the $S^{1}/\mathbb{Z}_{2}$ interval, one then finds a four-dimensional effective theory with an $E_{8} \times E_{8}$ gauge symmetry~\cite{Lukas:1997fg, Lukas:1998yy, Lukas:1998tt, Donagi:1998xe,Ovrut:2000bi}. The action of this theory matches that of the standard heterotic string, however the mass scales are governed by the Calabi--Yau radius and the orbifold length, which allow one to more easily obtain the unification gauge couplings and mass scale required in phenomenologically realistic GUT models. 
Other aspects of such theories, such as such as the spontaneous breaking of their supersymmetry \cite{Lukas:1999kt,Antoniadis:1997xk,Dudas:1997cd,Lukas:1997rb,Nilles:1998sx,Choi:1997cm}, the role of five-branes in the orbifold interval \cite{Lukas:1998hk,Lehners:2006ir,Carlevaro:2005bk,Gray:2003vw,Brandle:2001ts,Lima:2001nh,Grassi:2000fk} and the methods for stabilizing moduli \cite{Anderson:2010mh, Anderson:2011cza, Anderson:2011ty,Correia:2007sv}, have also been discussed in the literature.

Within this context there have been a large number of physically promising low-energy models constructed as top-down string models~\cite{Braun:2005nv,Braun:2005bw,Braun:2005ux,Bouchard:2005ag,Anderson:2009mh,Braun:2011ni,Anderson:2011ns,Anderson:2012yf,Anderson:2013xka,Nibbelink:2015ixa,Nibbelink:2015vha}. In particular, in \cite{Braun:2005bw,Braun:2005zv,Braun:2005nv, Braun:2006ae} it was shown that one can obtain the exact MSSM spectrum in Minkowski space by compactifying on a certain Schoen Calabi--Yau threefold equipped with a particular slope-stable $SU(4)$ gauge bundle. The resulting low-energy field content consists of three families of quarks and leptons with three right-handed neutrino chiral supermultiplets, one per family, and a Higgs-Higgs conjugate pair of chiral superfields~\cite{Braun:2005bw,Braun:2005zv,Braun:2005nv}. There are no vector-like pairs and no exotic fields. However, in addition to the gauge group $SU(3)_{C} \times SU(2)_{L} \times U(1)_{Y}$ of the MSSM, there is an extra gauged $U(1)_{B-L}$. Hence, this is referred to as the $B-L$ MSSM model~\cite{Marshall:2014kea,Marshall:2014cwa,Dumitru:2018jyb,Dumitru:2018nct,Dumitru:2019cgf}. A number of papers \cite{Deen:2016vyh,Ambroso:2009jd,Ovrut:2012wg,Ovrut:2014rba,Ovrut:2015uea,Barger:2008wn,FileviezPerez:2009gr,FileviezPerez:2012mj} discuss the physics of breaking this additional $U(1)_{B-L}$ symmetry above the electroweak scale.

For completeness, it is essential that there also exists a slope-stable gauge bundle on the hidden sector plane of the $B-L$ MSSM with the correct properties to ensure, for example, anomaly cancellation. The majority of previous work has focussed on constructing a realistic observable sector with the view that compatible hidden sectors would be found later. This view was supported by the results of 
\cite{Braun:2006ae}, which showed that hidden sector bundles are compatible with the Bogomolov inequality and so stable hidden sector bundles likely exist. The goal of this and several recent papers is to fill this gap by constructing these hidden sectors for the $B-L$ MSSM.

There have been many previous works on constructing and classifying heterotic line bundle backgrounds in both supersymmetric and non-supersymmetric theories -- see \cite{Anderson:2012yf,Honecker:2006qz,Nibbelink:2009sp,Blaszczyk:2010db,Anderson:2011ns,Anderson:2013xka,Nibbelink:2015ixa,Nibbelink:2015vha,Blaszczyk:2014qoa,Blaszczyk:2015zta,Braun:2017feb,Deen:2020dlf,Larfors:2020ugo,Larfors:2020weh,Otsuka:2020nsk} and references therein for a selection of these. In the context of the $B-L$ MSSM, viable hidden sector bundles were proposed in \cite{Braun:2013wr}, built from both single line bundles and direct sums of line bundles. Unfortunately, this was within the context of the weakly coupled heterotic string and so the values obtained for the unification scale and associated gauge couplings were not compatible with their expected values. In a previous work \cite{Ashmore:2020ocb}, we rectified this problem by moving to the strongly coupled heterotic string. We presented a hidden sector bundle built from a single line bundle $L$ via an induced rank-two bundle $L\oplus L^{-1}$ which, promisingly, satisfied all of the required ``vacuum'' constraints. We found a substantial region of K\"ahler moduli space in which the  $S^{1}/{\mathbb{Z}}_{2}$ orbifold scale was sufficiently large compared to the average Calabi--Yau radius, and where the effective strong coupling parameter was large enough, to obtain the correct values for the observable sector $SO(10)$ unification scale and gauge couplings. This hidden sector model, however, came with a caveat: although the results were computed only to order $\kappa_{11}^{4/3}$, the region of K\"ahler moduli space in which one is required to work to guarantee slope-stability generically leads to a large effective coupling parameter. Even though, by definition, we are working within the context of ``strongly coupled'' heterotic theory and, hence, higher-order corrections are not expected to be negligible, the large coupling parameter required in our case enhances this concern. That is, the linear approximation which is standard in heterotic M-theory theory and which was used in the $L \oplus L^{-1}$ bundle analysis became less trustworthy. 

This issue was explored in \cite{Ashmore:2020ocb} by computing the theory to the next order, that is, ${\cal{O}}(\kappa_{11}^{6/3})$. The results were reassuring since the ${\cal{O}}(\kappa_{11}^{6/3})$ corrections both strengthened and even improved upon the ${\cal{O}}(\kappa_{11}^{4/3})$ results. Be that as it may, this issue requires further investigation. With this in mind, an alternative construction of the hidden sector bundle from a single line bundle $L$ was preliminarily investigated in \cite{Ashmore:2020ocb}. Instead of the bundle $L\oplus L^{-1}$ discussed earlier, one now moves in K\"ahler moduli space so that the slope of $L^{-1}$ becomes negative. This significantly reduces the size of the effective coupling parameter and, hence, makes the ${\cal{O}}(\kappa_{11}^{4/3})$ expansion more trustworthy. To preserve supersymmetry, one must move in bundle moduli space to a rank-two hidden sector bundle $V$, defined by the extension $0 \rightarrow L^{-1} \rightarrow V \rightarrow L \rightarrow 0$. One must then show that the bundle $V$ is Bogomolov stable (a necessary but not sufficient condition for stability) and, second, that the extension is non-trivial; that is, that $H^{1}(L^{-2}) \neq 0$. A preliminary examination of this was carried out in \cite{Ashmore:2020ocb}. It was found that while the stability condition is relatively easy to satisfy, proving that the extension is non-trivial required a mathematical analysis beyond the scope of that paper. So, this approach also requires further investigation.
 
The purpose of the current work is to improve on these results. Our aim is to find new hidden sector bundles for which the linear approximation can be trusted while still satisfying all of the phenomenological and vacuum constraints laid out in \cite{Ashmore:2020ocb}. We focus on hidden sectors built from {\it two} line bundles, where the $U(1)\times U(1)$ structure group embeds as $U(1)\times U(1)\subset SU(3)\subset SU(3)\times E_6 \subset E_8$. We will show that there are a large number of examples which solve all of the phenomenological and vacuum constraints, as well as passing a number of non-trivial stability checks -- including Bogomolov stability. Deciding whether or not these candidate hidden sectors are viable is then reduced to checking whether certain $\op{Ext}^{1}$ groups are non-vanishing and then carrying out a more rigorous check of slope stability. Both of these analyses are beyond the scope of the present work -- we will to return to this in future publications. Our results indicate that it is very promising to build hidden sectors from two line bundles, and that using three or more line bundles would lead to even more examples.

In Section \ref{Physical Constraints}, we briefly review how the $B-L$ MSSM arises in heterotic $M$-theory, with a focus on the physical constraints we impose that are independent of the explicit construction of the hidden sector. In Section \ref{2 line bundles embedding}, we outline how one can construct a hidden sector bundle from multiple line bundles using the formalism of ``line bundle vectors''~\cite{Nibbelink:2016wms}. For specificity, we focus on embedding two line bundles into the $SU(3)$ factor of $SU(3)\times E_6 \subset E_8$, and compare the line bundle vector formalism with an explicit embedding via $U(1)$ subgroups of $SU(3)$. The anomaly condition and the formalism for computing the low-energy spectrum are then presented in this context. We discuss how to deform the corresponding Whitney sum bundle to an irreducible $SU(3)$ bundle, and give a number of necessary conditions for this non-Abelian bundle to be slope-stable. We then scan over possible choices of pairs of line bundles and reduce the question of whether or not the corresponding hidden sector bundles exist to the calculation of certain bundle-valued cohomologies. In Section \ref{sec:more_branches}, we extend this analysis to other branches in bundle moduli space and perform a scan over line bundles once more. We find a large number of candidate line bundles that could lead to viable hidden sectors. The appendices contain a discussion of anomaly cancellation, the linearized approximation for heterotic M-theory, the genus-one corrected Fayet--Iliopoulos terms for $U(1)$s, expressions for the gauge couplings, and a discussion of the subbundles of the extension bundle in each extension class.

\section{\texorpdfstring{$B-L$}{B-L} MSSM and Universal Constraints}\label{Physical Constraints}

The Ho\v{r}ava--Witten vacuum is an $S^{1}/\mathbb{Z}_{2}$ orbifold of M-theory with two ten-dimensional planes, one at each fixed point, separated by a one-dimensional interval. Each fixed plane, the observable and hidden sector planes respectively, has an $E_{8}$ gauge group. Heterotic M-theory is obtained by compactifying six of the ten remaining dimensions on a Calabi--Yau threefold. Heterotic M-theory compactification has been discussed widely in the literature -- see, for example, ~\cite{Braun:2004xv,Braun:2005bw,Braun:2005nv,Braun:2005zv}. We now  present a brief outline of the aspects of the $B-L$ MSSM vacuum of heterotic M-theory that are relevant for the present paper.

\subsection{\texorpdfstring{$B-L$}{B-L} MSSM Vacuum }\label{Physical Constraints1}

The first step in constructing the $B-L$ vacuum of heterotic M-theory is to compactify six of the ten dimensions on a particular Calabi--Yau manifold -- specifically, a Schoen threefold quotiented by a freely acting $\mathbb{Z}_3\times\mathbb{Z}_3$~\cite{MR923487,Braun:2004xv}. Secondly, on the Schoen threefold of the observable plane, one places a holomorphic vector bundle with structure group $SU(4)\subset E_8$. The connection on the $SU(4)$ bundle breaks the $E_8$ group down to 
\begin{equation}
  E_8 \to Spin(10) 
\end{equation}
in four dimensions, leading to a $Spin(10)$ ``grand unified'' group in the observable sector. This GUT group is then broken further at a scale $\langle M_{U}\rangle$ of order $10^{16} \text{ GeV}$ to the gauge group of the $B-L$ MSSM by turning on two Wilson lines, each associated with a different $\Z_3$ factor of the $\Z_3 \times \Z_3$ holonomy. This preserves the $N=1$ supersymmetry of the four-dimensional effective theory, but breaks the observable sector group to
\begin{equation}
  Spin(10) 
  \to 
  SU(3)_C \times SU(2)_L \times U(1)_Y \times U(1)_{B-L} \ ,
  \label{17}
\end{equation}
which is the gauge group of the MSSM with an additional gauged $U(1)_{B-L}$ symmetry. The spectrum of the $B-L$ MSSM is determined by the structure of the Schoen threefold~\cite{MR923487,Braun:2005zv} and the precise choice of $SU(4)$ bundle~\cite{Braun:2006ae}. More details can be found in \cite{Braun:2005nv,Braun:2005bw}. The spectrum contains exactly the three quark and lepton families of the  
MSSM, including three additional right-handed neutrino chiral multiplets, one per family. It also contains the conventional $H_{u}$ and $H_{d}$ Higgs doublet supermultiplets. There are no exotic fields or vector-like pairs.

The observable and hidden sectors are separated by a one-dimensional interval $S^{1}/\mathbb{Z}_{2}$. Within this interval, parallel to the orbifold planes, the theory allows for the existence of multiple five-branes.
In our previous paper \cite{Ashmore:2020ocb}, for simplicity, we assumed that all such branes coalesced into a single five-brane located near the hidden sector plane. We will make the same assumptions in the present work.
To preserve $N=1$ supersymmetry, this five-brane must wrap a holomorphic cycle in the Schoen threefold. This is equivalent to the topological class $W$ of the five-brane being ``effective''; that is,  
\begin{equation}
W_i \geq0\ ,\quad i=1,2,3\ .
\label{burt1}
\end{equation}
The numbers $W_i$ which characterize the five-brane class are defined in Appendix \ref{app:anomaly_cancellation}.

The remaining ingredient of the $B-L$ MSSM vacuum is the hidden sector. 
In previous work in \cite{Ashmore:2020ocb}, we constructed a hidden sector using a  single line bundle. However, the main goal of this present work is to construct slope-stable hidden sector bundles built from two line bundles. As we will show, this greatly increases the number of viable $B-L$ MSSM hidden sectors. This work will be presented in Sections 3 and 4.

An important aspect of the $B-L$ MSSM vacuum is the set of geometric moduli associated with it. The relevant moduli are the following. The Schoen threefold on which the Ho\v{r}ava--Witten theory is compactified is parametrized by the three real, positive K\"ahler moduli, $(a^1,a^2,a^3)$. These determine, for example, the volume of the Calabi--Yau threefold and, hence, are associated with the compactification scale. In addition, the $S^{1}/\mathbb{Z}_{2}$ interval between the observable and hidden sectors is parametrized by a single real modulus, $\hat{R}$. This determines the physical length, and hence the mass scale, of the interval. 
In \cite{Ashmore:2020ocb}, we derived a set of constraints on the values that $(a^1,a^2,a^3)$ and $\hat{R}$ must satisfy, thus restricting us to a specific region inside the positive K\"ahler cone. These constraints are a mixture of geometrical and phenomenological constraints, such as slope-stability of the observable sector $SU(4)$ bundle or ensuring the gauge couplings are compatible with experimental bounds. Some of these constraints can depend on the choice of hidden sector bundle, but only weakly through the effective five-brane class $W_{i}$. We call this set of constraints ``universal'', and present them in the next subsection. On the other hand, the constraints required to construct a consistent hidden sector bundle are independent of these universal physical constraints. These ``hidden sector bundle constraints'' will be discussed within the context of line bundle hidden sectors in
Sections \ref{2 line bundles embedding} and 4.
Our goal is to find hidden sector gauge bundles which are slope-stable inside the region of K\"ahler moduli space carved out by the universal set of constraints. More specifically, we will search for such bundles in the sub-region of this moduli space where the linear approximation to the vacuum is strictly valid -- that is, in the region of relatively small effective coupling. 

\subsection{Universal Physical Constraints}\label{Physical Constraints2}

The universal physical constraints, and the geometrical and phenomenological requirements leading to them, were discussed in detail in \cite{Ashmore:2020ocb}. In this subsection, we briefly review them since they impose important constraints on the physically allowed region of K\"ahler and $\hat{R}$ moduli space.

\begin{enumerate}
	
\item In order to preserve $N=1$ supersymmetry in the four-dimensional effective theory, the $SU(4)$ bundle must be both slope-stable and have vanishing slope~\cite{Braun:2005zv,Braun:2006ae}. As proven in detail in~\cite{Braun:2006ae}, the $SU(4)$ bundle in the observable sector of the $B-L$ MSSM is slope-stable, and, hence, admits a connection that satisfies the Hermitian Yang--Mills (HYM) equations, in the regions of the positive K\"ahler cone defined by
\begin{equation}  \label{51}
  \begin{gathered}
    \left(
      a^1
      < 
      a^2
      \leq 
      \sqrt{\tfrac{5}{2}} a^1
      \quad\text{and}\quad
      a^3
      <
      \frac{
        -(a^1)^2-3 a^1 a^2+ (a^2)^2
      }{
        6 a^1-6 a^2
      } 
    \right)
    \quad\text{or}  \\
    \left(
      \sqrt{\tfrac{5}{2}} a^1
      <
      a^2
      <
      2 a^1
      \quad\text{and}\quad
      \frac{
        2(a^2)^2-5 (a^1)^2
      }{
        30 a^1-12 a^2
      }
      <
      a^3
      <
      \frac{
        -(a^1)^2-3 a^1 a^2+ (a^2)^2
      }{
        6 a^1-6 a^2
      }
    \right) \ .
  \end{gathered}
\end{equation}

\item The squares of the ``unified'' gauge couplings in both the observable and hidden sectors must be positive definite.
As shown in Appendix \ref{app:gauge_couplings}, these conditions can be written as
\begin{align}
d_{ijk}a^ia^ja^k+3\frac{\epsilon_S^\prime \hat R}{V^{1/3}}\left(\tfrac{2}{3}a^1-\tfrac{1}{3}a^2+4a^3+(\tfrac{1}{2}-\lambda)^2W_ia^i\right)&>0 \ , \label{tree1}\\
d_{ijk}a^ia^ja^k-3\frac{\epsilon_S^\prime \hat R}{V^{1/3}}\left(\tfrac{2}{3}a^1-\tfrac{1}{3}a^2+4a^3+(1-(\tfrac{1}{2}+\lambda)^2)W_ia^i\right)&>0 \ ,
\label{eq:positive2}
\end{align}
where $z=\tfrac{1}{2}+\lambda$ gives the position of the five-brane in the interval. Following \cite{Ashmore:2020ocb,Ashmore:2020wwv}, in this paper we set the five-brane close to the hidden wall at $\lambda=0.49$. 
Note that for this choice, $(1-\tfrac{\lambda}{2})^2\sim {10}^{-4}$ and $(1-(\tfrac{1}{2}+\lambda))^2\sim {10}^{-2}$. Therefore, to a good approximation, the terms proportional to the five-brane charges $W_i$ drop out from the expressions above.

\item  The four-dimensional effective theory is derived by first compactifying on a Calabi--Yau threefold to give a five-dimensional theory, and then reducing further on the $S^{1}/{\mathbb{Z}}_{2}$ interval. For this to be consistent, we require that the length of the interval is sufficiently large compared to the average Calabi--Yau radius. This condition takes the form
\begin{equation}
\frac{\pi \rho {\Rhat} V^{-1/3}}{(vV)^{1/6} } > 1 \ .
\label{seb3}
\end{equation}

\item We want our top-down model to give reasonable four-dimensional physics, which leads to a number of ``phenomenological'' constraints. One such constraint is that the $Spin(10)$ grand unification scale, $\langle M_{U} \rangle$, and the associated unified gauge coupling in the observable sector, $\langle \alpha_{u}\rangle=\langle g^{(1)}\rangle^2/4\pi$, be consistent with phenomenologically acceptable values for these quantities. As discussed in \cite{Ashmore:2020ocb}, a reasonable choice for these quantities, which we will use in this paper, is
\begin{equation}
\langle M_{U}\rangle=3.15 \times 10^{16}~\text{GeV} \ ,\qquad \langle \alpha_{u} \rangle = \frac{1}{26.46} \ .
\label{jack1}
\end{equation}

\item  The reduction on the orbifold interval uses a linearized approximation to the five-dimensional BPS solution of heterotic M-theory~\cite{Lukas:1998tt}. In Appendix B we show that this requirement takes the form
\begin{equation}
 2\epsilon_S'\frac{\Rhat}{V^{1/3}}
  \left|
    \beta_i^{(0)} \big(z-\tfrac{1}{2}\big)
    -\tfrac{1}{2}W_i(\tfrac{1}{2}-\lambda)^2
  \right|
  \ll 
  | d_{ijk} a^j a^k |
  \ , \quad z \in [0, \lambda + \tfrac{1}{2}] \label{83} \ ,
  \end{equation}
\begin{equation}
2\epsilon_S'\frac{\Rhat}{V^{1/3}}
  \left|
    (\beta_i^{(0)}+W_i)
    \big(z-\tfrac{1}{2}\big)
    -\tfrac{1}{2}W_i(\tfrac{1}{2}+\lambda)^2
  \right| 
  \ll 
  | d_{ijk} a^j a^k |
  \ , \quad z \in [\lambda + \tfrac{1}{2},1] \ .
  \label{84}
\end{equation}
The $d_{ijk}$ are the intersection numbers for the specific Schoen threefold -- see \eqref{4new}. With the five-brane placed near to the hidden wall, $\lambda\approx 0.5$, these two conditions simplify to
\begin{equation}
\label{eq:linearizedCons}
d_{ijk} a^j a^k\gg \epsilon_S'\frac{\Rhat}{V^{1/3}}\beta_i^{(0)}\ .
\end{equation}
Specifically, \eqref{eq:linearizedCons} is valid for the choice choice of $\lambda=0.49$ used in\cite{Ashmore:2020ocb,Ashmore:2020wwv} and to simplify \eqref{tree1} and \eqref{eq:positive2} above.

\end{enumerate}

\begin{figure}[t]
   \centering
\includegraphics[width=0.5\textwidth]{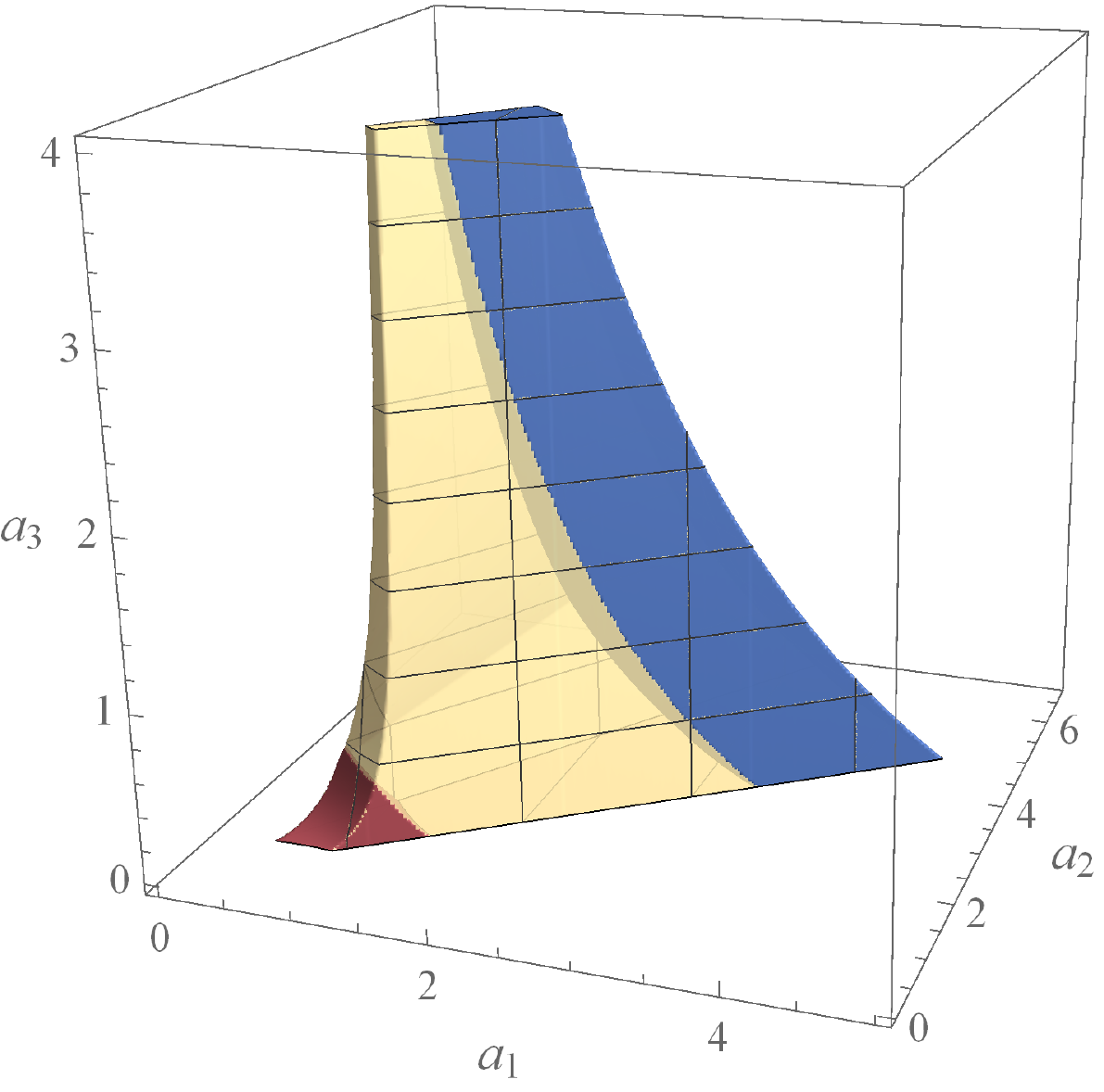}
\caption {Working in unity gauge, we show the entire region of K\"ahler moduli space where the universal conditions \eqref{51}--\eqref{jack1} are satisfied. To a good approximation, this subspace is independent of the
hidden sector bundle. In the ``red'' region, the theory is very strongly coupled. Within this region, the 
effective expansion parameter $\epsilon_S^{\text{eff}}$ is greater than 1. The ``blue'' region represents 
the region of K\"ahler moduli space in which, in addition to all the universal conditions being satisfied, the linearization constraint \eqref{eq:linearizedCons} is also valid -- that is, where $\epsilon_S^{\text{eff}} \lesssim 0.2$. The intermediate region, where $0.2<\epsilon_{S}^\text{eff}<1$, is displayed in ``yellow''. The main focus of this paper is to find a hidden sector bundle whose region of stability in moduli space intersects the ``blue'' region.}
\label{fig:ModuliSolution_orange}
\end{figure}

As discussed in \cite{Ashmore:2020ocb,Ovrut:2018qog}, the actual strong coupling expansion parameter used in the linearized approximation to the heterotic M-theory vacuum is given by
\begin{equation}
\epsilon_{S}^{\rm eff}=\frac{\epsilon_{S}' \hat{R}}{V} \ .
\label{tree2}
\end{equation}
We point out that this is, up to a constant factor of order one, precisely the strong coupling parameter presented in equation (1.3) of \cite{Banks:1996ss}. Since $V=\frac{1}{6}d_{ijk}a^{i}a^{j}a^{k}$, it follows that $\epsilon_{S}^\text{eff}$ is both K\"ahler and $\hat{R}$ moduli dependent. We find that the linearization conditions \eqref{84} and \eqref{eq:linearizedCons}, as well as the the linearized approximation to the square of the gauge couplings, given in \eqref{tree1} and \eqref{eq:positive2}, are satisfied only in the ``weak coupling'' region of moduli space where $\epsilon_{S}^{\text{eff}} \lesssim 0.2$. On the other hand, the linearization conditions and, hence, the linearized approximation clearly breaks down in the ``strong coupling'' regime of moduli space where $\epsilon_{S}^{\text{eff}} \gtrsim 1$.

Note that these all of the above constraints are invariant under the rescaling
\begin{equation}
a^{i} \rightarrow \mu a^{i}\ ,\qquad  \epsilon'_{S} \Rhat \rightarrow \mu^{3} \epsilon'_{S}\Rhat \ , 
\label{er4}
\end{equation}
for $\mu\in \mathbb{R}^+$ and so one can absorb the coupling $\epsilon'_{S}{\Rhat}/V^{1/3}$ into the definition of the moduli. This is equivalent to setting
\begin{equation}
\frac{\epsilon'_{S}{\Rhat}}{V^{1/3}}=1 \ .
\label{er33}
\end{equation}
We refer to this choice as ``unity'' gauge~\cite{Braun:2013wr,Ashmore:2020ocb}. Making use of this simplification, we can find the subregion of the K\"ahler cone in which the universal constraints \eqref{51}--\eqref{jack1} are all satisfied. We display this as the entire colored region in Figure \ref{fig:ModuliSolution_orange}. In \cite{Ashmore:2020ocb,Ashmore:2020wwv} we were forced to work in a very strongly coupled regime since we could not preserve supersymmetry in the hidden sector unless the 
genus-one corrected slope of the line bundles vanished. Such a cancellation is possible only if the effective expansion parameter 
$\epsilon_S^{\text{eff}}$ was larger than 1. It follows that the linearization constraint \eqref{eq:linearizedCons} is not satisfied and, hence, the linearized approximation to various quantities, such as the gauge couplings in \eqref{tree1} and \eqref{eq:positive2}, are uncertain. The region of K\"ahler moduli space in which $\epsilon_S^{\text{eff}}>1$ is shown in ``red'' in Figure \ref{fig:ModuliSolution_orange}. We want to move away from this very strongly coupled regime and find hidden sector bundles for which $\epsilon_{S}^\text{eff} \lesssim 0.2$ and, hence, the linearization constraint \eqref{eq:linearizedCons} and linearized approximations {\it are} valid. The region of K\"ahler moduli space in which this holds is shown in ``blue'' in Figure \ref{fig:ModuliSolution_orange}. Finally, the ``intermediate'' region, where $0.2<\epsilon_{S}^\text{eff}<1$, is displayed in ``yellow''. The focus of this paper is to find hidden sector bundles which can be stable when the K\"ahler moduli $(a^1,a^2,a^3)$ are in the blue region -- and so satisfy all of the universal constraints as well as the linearization condition.

\section{Two Line Bundle Embedding}\label{2 line bundles embedding}

We showed in \cite{Ashmore:2020ocb,Ashmore:2020wwv}  that vacuum configurations with hidden sectors built from a single line bundle require that the genus-one corrected slope of the line bundle vanishes. It seems that such a configuration always pushes us into a strongly coupled regime in which the accuracy of the linear approximation used to derive the effective four-dimensional theory is uncertain. Here, we begin with a general overview of constructing Abelian line bundle backgrounds, before specializing our discussion to hidden sectors built from {\it two} line bundles embedded into the hidden $E_8$ gauge group. As an example of how this works, we will then focus on embeddings which lead to the breaking pattern $E_8\to E_6\times U(1)\times U(1)$. 

\subsection{Line Bundle Embeddings}\label{sec:embeddings}

A particularly simple set of hidden sector bundles are those constructed from line bundles. These are defined by a set of line bundles and the embedding of their corresponding $\Uni 1$ groups into $\Ex 8$. Given a set of line bundles, there are multiple inequivalent ways to embed their Abelian gauge connections into the ten-dimensional hidden $\Ex 8$ connection. A particularly useful formalism for describing these embeddings is using ``line bundle vectors''. Following \cite{Nibbelink:2016wms}, the Abelian gauge connections can be embedded in the hidden sector $\Ex 8$ by expanding the curvature $F_{E_{8}}$ as
\begin{equation}
	\frac{F_{E_{8}}}{2\pi}=\frac{1}{v^{1/3}}\omega_{i}H_{i}\ ,\label{eq:flux-1}
\end{equation}
where the coefficients $H_{i}$ are matrices valued in the Lie algebra of $\Ex 8$. As in \cite{Ashmore:2020ocb,Ashmore:2020wwv}, the $\omega_{i}$ are the three harmonic $(1,1)$-forms that span the $H^{1,1}(X,\mathbb{C})$ cohomology on the Schoen threefold $X$, with their intersection numbers $d_{ijk}$ given by \eqref{4new}. Since the background is Abelian, one can expand the coefficients as
\begin{equation}
	H_{i}=V_{i}^{I}H_{I}\ ,
\end{equation}
where $I=1,\ldots,8$ runs over the Cartan subalgebra of the hidden $\Ex 8$. Here the $H_{I}$ denote the Cartan generators of the $\SO{16}\subset\Ex 8$ maximal subgroup, normalised so that
\begin{equation}\label{gen_norm}
	\tr H_{I}H_{J}=2\,\delta_{IJ}\ ,
\end{equation}
where the trace is $1/30$ of the trace over the $\rep{248}$ of $\Ex 8$ or, equivalently, taken in the fundamental $\rep{16}$ representation of $\SO{16}$.\footnote{This agrees with the normalisation in \cite{Nibbelink:2016wms} after noting that their trace is taken in the fundamental of an SU group rather than SO. See \cite[Appendix A]{Blaszczyk:2015zta} for more details.} The eight-component vectors $\boldsymbol{V}_{i}=V_{i}^{I}$ are known as \emph{line bundle vectors}. Given a choice of Cartan generators, an Abelian hidden sector bundle is completely specified by a choice of three line bundles vectors $\boldsymbol{V}_{i}$, $i=1,2,3$. For example, $\boldsymbol{V}_{1}$ then encodes how much $\omega_{1}$ contributes to the curvature $F_{E_8}$. This is somewhat abstract at the moment, but we will see how this works with an explicit example later.

As noted in \cite{Nibbelink:2015ixa}, the flux $F_{E_{8}}$ has to be quantized when evaluated on a string state and integrated over any curve dual to a divisor defined by a sum of the $\omega_{i}$. Since the string states are characterized by weight vectors that lie on the $\Ex 8$ root lattice $\Lambda$, the line bundle vectors must also lie on the root lattice, that is $\boldsymbol{V}_{i}\in\Lambda.$ Following the conventions of \cite{Feger:2012bs,Feger:2019tvk}, the $\Ex 8$ root lattice is given by the set of points $\Lambda\in\bR^{8}$ such that all eight coordinates are integers or half-integers (but not a mix of the two), and the coordinates sum to an even integer. This constrains the form of the line bundle vectors and ensures that the curvature of the resulting hidden sector bundle obeys flux quantisation.

The second Chern character of the hidden sector bundle $\mathcal{V}^{(2)}$ constructed from the line bundles is given by
\begin{equation}
	\text{ch}_{2}(\mathcal{V}^{(2)})=\frac{1}{16\pi^{2}}\tr F_{E_{8}}\wedge F_{E_{8}}=\tfrac{1}{2}(\boldsymbol{V}_{i}\cdot\boldsymbol{V}_{j})\,\frac{1}{v^{2/3}}\omega_{i}\wedge\omega_{j}\ ,\label{eq:chern2-1}
\end{equation}
where $\boldsymbol{V}_{i}\cdot\boldsymbol{V}_{j}=V_{i}^{I}V_{j}^{I}$ is the Euclidean scalar product between the $i^{\text{th}}$ and $j^{\text{th}}$ line bundle vectors. Since $c_{1}(\mathcal{V})=0$ -- following from the fact that the generators $H_{I}$ are traceless -- the second Chern class is given by $c_{2}(\mathcal{V}^{(2)})=-\text{ch}_{2}(\mathcal{V}^{(2)})$. Wedging with $\omega_{i}$ and integrating over $X$, one finds that the second Chern numbers of $\mathcal{V}^{(2)}$ are
\begin{equation}
c_{2,i}(\mathcal{V}^{(2)})\equiv\frac{1}{v^{1/3}}\int_X c_2(\mathcal{V}^{(2)})\wedge\omega_i	 = -\tfrac{1}{2}d_{ijk}(\boldsymbol{V}_{j}\cdot\boldsymbol{V}_{k})\ .
\end{equation}

The unbroken gauge group $G$ in four dimensions is given by the commutant of the structure group of $\mathcal{V}^{(2)}$ with the ten-dimensional gauge group. The \emph{non-Abelian} part of $G$ can be computed by finding all roots\footnote{The 240 roots are given by vectors $\boldsymbol{r}$ that lie on the root lattice with length squared equal to 2; $\{\boldsymbol{r}\in\Lambda\;|\;\boldsymbol{r}\cdot \boldsymbol{r}=2\}$.} $\boldsymbol{r}$ of $\ex 8$ that are orthogonal to all of the line bundle vectors:
\begin{equation}
	H_{i}(\boldsymbol{r})\equiv\boldsymbol{V}_{i}\cdot \boldsymbol{r}=0\ ,\quad\text{for all }i=1,\ldots,h^{1,1}\ .\label{eq:unbroken_condition}
\end{equation}
This ensures that the components of the $\Ex 8$ connection that form the four-dimensional connection are uncharged with respect to $G$. Since we are considering bundles with an Abelian structure group, there may also be $\Uni 1$ factors in $G$ (since they commute with themselves).

One can also calculate the chiral part of the matter spectrum in the resulting four-dimensional theory by computing the Euler characteristic $\chi$ for the bundles in which the various matter fields transform. This can be done using the multiplicity operator $\mathcal{N}$~\cite{Nibbelink:2007rd}. More precisely, the Euler characteristic $\chi(V)$ for a bundle $V$ whose sections transform in the representation $\rep R$ is calculated by $\mathcal{N}(\boldsymbol{r})$, where $\boldsymbol{r}$ is the $\ex 8$ root corresponding to the highest weight of $\rep R$. That is, given a decomposition of the $\rep{248}$ into representations $\rep R$, each $\rep R$ is characterized by some highest weight, which corresponds to some root of $\ex 8$ (since the roots are the weights of the $\rep{248}$). To compute $\chi(V)$, one simply finds the $\ex 8$ root $\boldsymbol{r}$ that the highest weight of $\rep R$ corresponds to and then evaluates $\mathcal{N}(\boldsymbol{r})$ as
\begin{equation}
	\chi(V)=\mathcal{N}(\boldsymbol{r})\equiv\tfrac{1}{12}c_{2,i}(X)H_{i}(\boldsymbol{r})+\tfrac{1}{6}d_{ijk}H_{i}(\boldsymbol{r})H_{j}(\boldsymbol{r})H_{k}(\boldsymbol{r})\ ,\label{eq:mult-1}
\end{equation}
where the second Chern numbers of the Schoen threefold are $c_{2,i}(X)=(4,4,0)_i$. Note that we will often abuse notation and write $\chi(\rep R)$ for the Euler characteristic of the bundle $V$ transforming in the $\rep R$ representation. With these conventions, a left chiral fermion zero-mode in four dimensions has $\mathcal{N}(\boldsymbol{r})<0$.

\subsection{Embedding Constraints for Two Line Bundles}

As mentioned above, since our Schoen threefold has $h^{1,1}=3$, we need to specify three line bundle vectors $\boldsymbol{V}_{i}$. To be concrete, we now consider the explicit example of a line bundle background that breaks $\Ex 8$ to $\Ex 6\times\Uni 1\times\Uni 1$. We will describe this bundle using both the line bundle vector description above and a more standard description. 

It is useful to decompose the $\boldsymbol{V}_{i}$ into a set of line bundle data $(m^{i},n^{i})$ and a set of linearly independent, eight-component basis vectors $(\boldsymbol{t}_{1},\boldsymbol{t}_{2})$. The line bundle vectors can then be written as
\begin{equation}\label{line_vecs}
	\boldsymbol{V}_{i}=m^{i}\boldsymbol{t}_{1}+n^{i}\boldsymbol{t}_{2}\ .
\end{equation}
To match with our previous conventions in the case of the single line bundle, we take the generators of the $\Uni 1\times\Uni 1$ structure group to be $(-\boldsymbol{t}_{1},-\boldsymbol{t}_{2})$. One then finds that matter fields transform according to
\begin{equation}\label{eq:bundle_def}
	\rep 1_{-1,0}\sim\mathcal{O}_{X}(m^{1},m^{2},m^{3})=L_{1}\ ,\qquad\rep 1_{0,-1}\sim\mathcal{O}_{X}(n^{1},n^{2},n^{3})=L_{2}\ ,
\end{equation}
From this we see that $(m^{i},n^{i})$ specify the line bundles, while $(\boldsymbol{t}_{1},\boldsymbol{t}_{2})$ give the embedding of $\Uni 1\times\Uni 1$ into the hidden sector $\Ex 8$. If we want to consider the embedding of a single $\Uni 1$ into $\Ex 8$, one takes $\boldsymbol{t}_{2}=\boldsymbol{0}$, while more $\Uni 1$s could be embedded by including more basis vectors.\footnote{This should be compared with the discussion in \cite[Section 7.2]{Nibbelink:2016wms}, where they give an example of a $\Uni 1 \times \Uni 1$ bundle with three line bundle vectors and a single relation between them, implying that they can be written in a basis with two linearly independent generators.} More details on the generators and our conventions can be found in Appendix \ref{sec:Conventions}. The eight-component basis vectors are taken to be
\begin{equation}\label{eq:u1_vectors}
	\boldsymbol{t}_{1}=(0,0,0,0,0,-1,1,0)\ ,\qquad\boldsymbol{t}_{2}=(0,0,0,0,0,-1,-1,-2)\ ,
\end{equation}
where the two $\Uni 1$ groups are generated by $(-\boldsymbol{t}_{1},-\boldsymbol{t}_{2})$. Note that this choice obeys flux quantisation when the entries of $m^{i}$ and $n^{i}$ are integers. Note also that one must have $m^{1}+m^{2}\mod3=0$ for equivariance of the line bundle $L_{1}$, with the same condition for the $n^{i}$ as well.

It is easy to see which simple roots of $\Ex 8$ are broken by this choice – one simply takes the inner product of each $\boldsymbol{V}_{i}$ with the simple roots in (\ref{eq:roots}). In particular, $\boldsymbol{t}_{1}$ breaks a combination of $\alpha_{6}$ and $\alpha_{7}$, while $\boldsymbol{t}_{2}$ breaks $\alpha_{6}$ alone. Together, they break both $\alpha_{6}$ and $\alpha_{7}$,\footnote{One can also see this by transforming the basis vectors to the $\omega$-basis, where they are given by $(0,0,0,0,0,-1,2,0)$ and $(0,0,0,0,0,-1,0,0)$.} suggesting that the unbroken gauge group will be $\Ex 6$. One can check this explicitly using (\ref{eq:unbroken_condition}) from which one sees that 72 roots of $\Ex 8$ are annihilated by the line bundle vectors, which then form the 72 roots of the unbroken $\Ex 6$ which commutes with $\Uni 1\times\Uni 1$ inside $\Ex 8$. Using the \emph{Mathematica }package LieART~\cite{Feger:2012bs,Feger:2019tvk}, one can then find the decomposition of the adjoint representation of $\Ex 8$:
\begin{equation}
	\label{eq:decompE8SU3big}
	\begin{split}
		\rep{248}&=\rep{78}_{0,0}+2\times\rep{1}_{0,0}+\rep{1}_{2,0}+\rep{1}_{-2,0}+\rep{1}_{1,3}+\rep{1}_{-1,3}+
		\rep{1}_{1,-3}+\rep{1}_{-1,-3}\\
		&\eqspace+\rep{27}_{1,-1}+\rep{27}_{-1,-1}+\repb{27}_{1,1}+\repb{27}_{-1,1}+\rep{27}_{0,2}+\repb{27}_{0,-2}\ .\\
	\end{split}
\end{equation}

This particular breaking pattern can be obtained in a more conventional manner by first breaking $E_8\rightarrow E_6\times SU(3)$, under which the adjoint representation of $E_8$ decomposes as
\begin{equation}
\label{eq:E8SU3first}
\bf{248}=(\rep{78},\rep{1})+(\rep{1},\rep{8})+(\rep{27},\repb{3})+(\repb{27},\rep{3})\ .
\end{equation}
Breaking $SU(3)$ further to $SU(2)\times U(1)$, the $SU(3)$ representations that appear above decompose as
\begin{equation}
\label{eq:decompSU3}
\begin{split}
\rep{3}&=\rep{2}_{1}+\rep{1}_{-2}\ ,\\
\repb{3}&=\rep{2}_{-1}+ \rep{1}_{2}\ ,\\
\rep{8}&=\rep{3}_{0}+\rep{2}_{3}+\rep{2}_{-3}+\rep{1}_0\ .\\
\end{split}
\end{equation}
Finally, we break the $SU(2)$ further to $U(1)$, with the $SU(2)$ representations above decomposing as
\begin{align}
\rep{2}&=\rep{1}_{1}+\rep{1}_{-1}\ ,\\
\rep{3}&=\rep{1}_2+\rep{1}_0+\rep{1}_{-2}\ .
\end{align}
Putting this together, one sees that the adjoint representation of $E_8$ decomposes under this $E_6\times U(1)\times U(1)$ exactly as in \eqref{eq:decompE8SU3big} above. We thus have two equivalent descriptions of the embedding of this $U(1)\times U(1)$ in $E_8$, either via the line bundle vectors defined by \eqref{eq:u1_vectors} or via a chain of subgroups starting from the maximal subgroup $E_6\times SU(3)$.

We can find the $E_8$ connection that corresponds to this breaking pattern by first building the $SU(3)$ connection induced by the breaking $SU(3)\rightarrow S(U(1)\times U(1)\times U(1))\simeq U(1)\times U(1)$ and two $U(1)$ connections, $A^{(1)}_{U(1)}$ and $A^{(2)}_{U(1)}$. The two $U(1)$s embed into $SU(3)$ as
\begin{equation}
( \ee^{\ii\phi_1},\ee^{\ii\phi_2})\hookrightarrow  \begin{pmatrix}\ee^{-2\ii\phi_2}&0\\0&\ee^{\ii\phi_2}\begin{pmatrix}\ee^{\ii\phi_1}&0\\0&\ee^{-\ii\phi_1}\end{pmatrix}\end{pmatrix} \ .
\end{equation}
Using this, we can build a connection associated with a rank three bundle $V_{\rep 3}$ as
\begin{equation}
	\begin{split}
		\label{eq:SU3U1U1}
		A_{SU(3)} = \begin{pmatrix}-2A^{(2)}_{U(1)}&0&0\\0&A^{(1)}_{U(1)}+A^{(2)}_{U(1)}&0\\0&0&-A^{(1)}_{U(1)}+A^{(2)}_{U(1)}\end{pmatrix}\ ,\\
	\end{split}
\end{equation}
where $A^{(1)}_{U(1)}$ and $A^{(2)}_{U(1)}$ are the $L_1$ and $L_2$  line bundle connections respectively. The form of connection \eqref{eq:SU3U1U1} implies that the rank three bundle is the Whitney sum
\begin{equation}
\label{eq:defineBundle2}
\begin{split}
V_{\rep 3}&= L_2^{-2}\oplus\left((L_1\oplus L_1^{-1})\otimes L_2\right)
\\&= L_2^{-2}\oplus L_1L_2\oplus L_1^{-1}L_2\equiv\mathcal{F}\oplus\mathcal{K}\oplus\mathcal{E}\ ,
\end{split}
\end{equation}
where we have defined
\begin{equation}\label{eq:FKL}
\mathcal{F}=L_2^{-2}\ , \qquad \mathcal{K}=L_1 L_2\ , \qquad \mathcal{E}=L_1^{-1} L_2\ .
\end{equation}
From the form of the $SU(3)$ connection in \eqref{eq:SU3U1U1}, one can read off that the $U(1)$ connections associated with the line bundles $\mathcal{F}$, $\mathcal{K}$ and $\mathcal{E}$ are 
\begin{equation}
	A^{\mathcal{F}}_{U(1)}=-2A^{(2)}_{U(1)}\ ,\qquad A^{\mathcal{K}}_{U(1)}=A^{(1)}_{U(1)}+A^{(2)}_{U(1)}\ , \qquad A^{\mathcal{E}}_{U(1)}=-A^{(1)}_{U(1)}+A^{(2)}_{U(1)}
\end{equation}
respectively.  Note that $A^{\mathcal{E}}_{U(1)}+A^{\mathcal{F}}_{U(1)}+A^{\mathcal{K}}_{U(1)}=0$ and, therefore, that
\begin{equation}
c_1(\mathcal{F})+c_1(\mathcal{K})+c_1(\mathcal{E})=0\ ,
\end{equation}
which is simply the condition that $V_{\rep 3}$ is an $SU(3)$ bundle.

The DUY theorem \cite{Donaldson, UY} implies that there exists a connection $A_{SU(3)}$ that solves the HYM equation if $V_{\rep 3}=\mathcal{F}\oplus\mathcal{K}\oplus\mathcal{E}$ is poly-stable. Since the slope of any $SU(N)$ bundle such as $V_{\rep 3}$ vanishes, it follows that $V_{\rep 3}$ can be poly-stable only if the slopes of each of its subbundles vanish as well. Hence, the hidden sector bundle $V_{\rep 3}$ given in 
\eqref{eq:defineBundle2} will be slope poly-stable if
\begin{equation}
\mu(\mathcal{F})=\mu(\mathcal{K})=0 \ .
\end{equation}
Note that the slope of $\mathcal{E}$ vanishes automatically if the slopes of $\mathcal{F}$ and $\mathcal{K}$ do, so the condition above is sufficient.

The $SU(3)$ connection in \eqref{eq:SU3U1U1} embeds further into an $E_8$ connection such that it commutes with $E_6$ as in \eqref{eq:E8SU3first}. 
The embedding of the line bundle connections into the hidden $E_8$ is then given by
\begin{equation}\label{eq:E8_connection}
(A^{(1)}_{U(1)},A^{(2)}_{U(1)})\hookrightarrow A_{E_8}=A^{(1)}_{U(1)}Q_1+A^{(2)}_{U(1)}Q_2\ ,
\end{equation}
where $Q_1$ and $Q_2$ are elements of the $E_8$ algebra whose traces obey
\begin{equation}\label{eq:traces}
	\tfrac{1}{4}\tr Q_1^2=1\ , \qquad \tfrac{1}{4}\tr Q_2^2=3\ , \qquad \tr Q_1Q_2=0\ ,
\end{equation}
which can also be read off from the decomposition in \eqref{eq:decompE8SU3big}. We see that $Q_1$ and $Q_2$ contain the charges associated with each of the two $U(1)$ subgroups.

We can also see this from the line bundle vector description as follows. Since the curvature can be expanded in the Cartan generators as in \eqref{eq:flux-1}, we can write
\begin{equation}
	F_{E_8} = \frac{2\pi}{v^{1/3}}\omega_i H_i = \frac{2\pi}{v^{1/3}} \omega_i (m^i t^I_1 + n^i t^I_2)H_I = (F_{U(1)}^{(1)} t^I_1 + F_{U(1)}^{(2)} t^I_2)H_I\ ,
\end{equation}
where we have identified $F_{U(1)}^{(1)}=2\pi v^{-1/3} m^i\omega_i$ as the curvature of the line bundle $L_1=\mathcal{O}_X(m^1,m^2,m^3)$ defined in \eqref{eq:bundle_def}, and similarly for $L_2$. Comparing with \eqref{eq:E8_connection}, we read off that $Q_1 = t_1^I H_I$ and $Q_2 = t_2^I H_I$. It is then simple to repeat the calculation of the traces to give
\begin{equation}\label{eq:traces_line}
	\tfrac14\tr Q_1^2 = \tfrac12\boldsymbol{t}_1\cdot \boldsymbol{t}_1 = 1 \ , \qquad \tfrac14\tr Q_2^2 = \tfrac12\boldsymbol{t}_2\cdot \boldsymbol{t}_2 = 3\ , \qquad \tfrac14\tr Q_1 Q_2 = \tfrac12\boldsymbol{t}_1\cdot \boldsymbol{t}_2 = 0 \ ,
\end{equation}
where we have used the normalisation of the generators in \eqref{gen_norm}. We see that this agrees with \eqref{eq:traces} upon inserting the definitions from \eqref{line_vecs} and \eqref{eq:u1_vectors}.

\subsection*{Anomaly Condition}

The second Chern character of the hidden sector bundle is
\begin{equation}
\op{ch}_{2}(\mathcal{V}^{(2)})=\frac{1}{16\pi^{2}}\tr F_{E_8}\wedge F_{E_8},
\end{equation}
where $F_{E_8}$ is the curvature of the $E_8$ connection induced from the two line bundle connections in \eqref{eq:E8_connection}.
Expanding out, we find
\begin{equation}
\op{ch}_{2}(\mathcal{V}^{(2)})=\frac{1}{16\pi^{2}}\left(\tr Q_1^2F_{U(1)}^{(1)}\wedge F_{U(1)}^{(1)}+\tr Q_2^2F_{U(1)}^{(2)}\wedge F_{U(1)}^{(2)}+
2\tr Q_1Q_2F_{U(1)}^{(1)}\wedge F_{U(1)}^{(2)}\right) \ ,
\end{equation}
where the $F_{U(1)}^{(i)}$ are given in terms of the first Chern class of each line bundle as $ c_{1}^i=F^{(i)}_{U(1)}/2\pi$. Again denoting the $L_{1}$ and $L_{2}$ line bundles by 
\begin{equation}\label{eq:L1_L2_def}
L_1=\mathcal{O}_X(m^1,m^2,m^3)\ , \qquad L_2=\mathcal{O}_X(n^1,n^2,n^3)\ ,
\end{equation}
the second Chern character of $\mathcal{V}^{(2)}$ is then simply
\begin{equation}
\begin{split}\label{ch2}
\op{ch}_{2}(\mathcal V^{(2)})&=\,c_{1}(L_1)\wedge c_{1}(L_1)+3c_{1}(L_2)\wedge c_{1}(L_2)\\
&=\frac{1}{v^{2/3}}(m^1\omega_{1}+m^2\omega_{2}+m^3\omega_{3})^{2}+\frac{3}{v^{2/3}}(n^1\omega_{1}+n^2\omega_{2}+n^3\omega_{3})^{2}\ ,
\end{split}
\end{equation}
where we used the trace relations from \eqref{eq:traces}. It then follows from the discussion in Appendix \ref{app:anomaly_cancellation} that the anomaly condition is given by
\begin{equation}
\label{eq:anomaly_modified}
W_i=\left(\tfrac{4}{3},\tfrac{7}{3},-4  \right)_i+d_{ijk}m^jm^k+3d_{ijk}n^jn^k\ .
\end{equation}
Furthermore, to be consistent with $N=1$ supersymmetry, $W_{i}$ must satisfy the constraint given in \eqref{burt1}; that is, $W_i \geq 0$ for $i=1,2,3$.

Again we can compare this with the line bundle vector formalism. We have already given the second Chern character in \eqref{eq:chern2-1}. Expanding this out and using the trace relations in \eqref{eq:traces_line}, one can check that it reproduces \eqref{ch2} above. Furthermore, the anomaly cancellation condition can then be written as
\begin{equation}
	W_{i}=(\tfrac{4}{3},\tfrac{7}{3},-4)_{i}+\tfrac{1}{2}d_{ijk}(\boldsymbol{V}_{j}\cdot\boldsymbol{V}_{k})\ ,
\end{equation}
which agrees with \eqref{eq:anomaly_modified} upon using the definitions from \eqref{line_vecs} and \eqref{eq:u1_vectors}.

\subsection*{Low-Energy Fields}

As usual, low-energy matter superfields fields arising from the decomposition in eq.~\eqref{eq:decompE8SU3big} are associated with bundle-valued cohomologies on the Calabi--Yau threefold. Using the identification \eqref{eq:FKL}, we find that the $E_6$ singlets with non-zero $U(1)$ charges are associated with
\begin{align}
	\rep 1_{0,2}&\sim H^\bullet(X,\mathcal{F})=H^\bullet(X,L_2^{-2})\ ,\\
	 \rep 1_{-1,-1}&\sim H^\bullet(X,\mathcal{K})=H^\bullet(X,L_1 L_2)\ ,\\
	 \rep 1_{1,-1}&\sim H^\bullet(X,\mathcal{E})=H^\bullet(X,L_1^{-1}L_2)\ .
\end{align}
To find the cohomologies for the other representations in the decomposition, we can either use the fact that all $SU(3)$ representations can be obtained from the fundamental $\rep 3$ and its conjugate $\repb 3$, and then use the decomposition of the $\rep3$ in terms of $U(1)\times U(1)$, or we can just count the charges in \eqref{eq:decompE8SU3big} and use the above identifications to find the corresponding cohomology. 
The representations we obtain for the $E_6\times U(1)\times U(1)$ low-energy group, as well as their corresponding cohomologies, are shown in Table \ref{tab:4thTable}. Note that each representation $\rep R$ in Table \ref{tab:4thTable} has an associated cohomology of the form
\begin{equation}
H^\bullet(X,L_{\rep R})=H^\bullet(X,L_1^{-q_{\rep R}}\otimes L_2^{-p_{\rep R}})\ ,
\end{equation}
where $q_{\rep R}$ and $p_{\rep R}$ are the charges of $\rep R$ for each of the two $U(1)$ groups.

Let us now consider the low-energy matter spectrum. For fields in the $\rep R$ representation, with associated line bundle $L_{\rep R}$, the Euler characteristic $\chi(L_{\rep R})$ counts the chiral asymmetry. For a line bundle of the form $L_{\rep R}=\mathcal{O}_{X}(l^1_{\rep R},l^2_{\rep R},l^3_{\rep R})$, the Euler characteristic is given by
\begin{equation}
\chi(L_{\rep R})=\sum_{i=0}^{3}(-1)^ih^i(X,L_{\rep R})=\int_X \op{ch}(L_{\rep R})\wedge \op{Td}(X)\ ,
\label{green1}
\end{equation}
where  $\op{ch}(L_{\rep R})$ is the Chern character of $L_{\rep R}$ and $\op{Td}(X)$ is the Todd class of the tangent bundle of $X$. On the Schoen manifold we are considering, this simplifies to 
\begin{equation}
\label{eq:Euler_ch_li}
\chi(L_{\rep R})=\tfrac{1}{3}(l^1_{\rep R}+l^2_{\rep R})+\tfrac{1}{6}d_{ijk}l^i_{\rep R}l^j_{\rep R}l^k_{\rep R}\ .
\end{equation}
Using the intersection numbers $d_{ijk}$ given in \eqref{4new}, this expression becomes
\begin{equation}
\chi(L_{\rep R})=\tfrac{1}{6}\left(l^1_{\rep R}l^2_{\rep R}(l^1_{\rep R}+l^2_{\rep R}+6l^3_{\rep R})+2l^1_{\rep R}+2l^2_{\rep R}\right)\ .
\label{hope2}
\end{equation}
The numbers $l_{\rep R}^i$ characterizing the line bundle $L_{\rep R}$ depend on the low-energy representation $\rep R$. For our line bundle embedding and a representation $\rep R$ with $U(1)$ charges $q_{\rep R}$ and $p_{\rep R}$, the Euler characteristic is given by $\chi(L_{\rep R})=\chi({L_1^{-q_{\rep R}}L_2^{-p_{\rep R}}})$. Defining $L_1$ and $L_2$ as in \eqref{eq:L1_L2_def}, one finds that 
\begin{equation}
 l_{\rep R}^i=-q_{\rep R}m^i-p_{\rep R}n^i\ , \quad i=1,2,3\ .
\end{equation}
It is then simple to evaluate the Euler characteristic for each representation by substituting the
values of $l_{\rep R}^i$ into \eqref{hope2}.
For example, for the line bundle $L_1^{-1}L_2^{-3}$ associated with the representation $\rep 1_{1,3}$ in 
Table \ref{tab:4thTable}, the $l^i$ are
\begin{equation}
l^i=-m^i-3n^i\ .
\end{equation}
The Euler characteristic is then given by
\begin{equation}\label{chi}
\chi(\rep{1}_{1,3})\equiv\chi(L_1^{-1}L_2^{-3}) =\tfrac{1}{6}\left(l^1l^2(l^1+l^2+6l^3)+2l^1+2l^2\right)\ .
\end{equation}
It is straightforward to check that this agrees with the Euler characteristic as computed using the multiplicity operator in \eqref{eq:mult-1}. For example, the highest weight of $\rep{1}_{1,3}$ can be obtained by projecting the root $\boldsymbol{r}=(0,0,0,0,0,1,0,1)$ of $\mathfrak{e}_8$. Using this in \eqref{eq:mult-1}, one finds that $\mathcal{N}(\boldsymbol{r})$ evaluates to \eqref{chi}.

The matter fields associated with any given representation and cohomology in Table 1 are either 1) chiral superfields if the corresponding Euler characteristic $\chi <0$ or 2) anti-chiral superfields if $\chi >0$. For any given entry in Table 1, the sign of the Euler characteristic depends on the choice of line bundles $L_{1}$ and $L_{2}$. Hence, the same entry in the Table can correspond to either a chiral superfield or an anti-chiral superfield depending on the circumstances of the solution. Note, however, that if a representation, such as ${\underline{\bf 1}}_{1,3}$, corresponds to chiral superfields then the conjugate representation, such as ${\underline{\bf 1}}_{-1,-3}$, corresponds to anti-chiral superfields. With this in mind, in the last column of Table 1 we have assigned a specific symbol to the matter supermultiplet of each representation. Having done that, we note that each such superfield contains a complex scalar field component. In the following, it is convenient to abuse notation and denote each superfield and its complex scalar field component using the same symbol. Whether we are referring to the full supermultiplet or its scalar component will be clear from the context.

\begin{table}[t]
	\noindent \begin{centering}
		\begin{tabular}{clc}
			\toprule 
			$E_6\times U(1)\times U(1) $ & Cohomology & Field Name\tabularnewline
			\midrule
			\midrule 
			$\rep{1}_{1,3}$ & $H^{\bullet}(X,\mathcal{F}\otimes \mathcal{K}^*=L_1^{-1}L_2^{-3})$ &$C_1$\tabularnewline
			\midrule 
			$\rep{1}_{-1,-3}$ & $H^{\bullet}(X,\mathcal{F}^*\otimes \mathcal{K}=L_1L_2^{3})$& $\tilde C_1$\tabularnewline
			\midrule 
			$\rep{1}_{-1,3}$ & $H^{\bullet}(X,\mathcal{F}\otimes \mathcal{E}^*=L_1L_2^{-3})$ & $C_2$\tabularnewline
			\midrule 
			$\rep{1}_{1,-3}$ & $H^{\bullet}(X,\mathcal{F}^*\otimes \mathcal{E}=L_1^{-1}L_2^{3})$ &$\tilde C_2$\tabularnewline
			\midrule 
			$\rep{1}_{2,0}$ & $H^{\bullet}(X,\mathcal{K}^*\otimes \mathcal{E}=L_1^{-2})$ &$C_3$\tabularnewline
			\midrule 
			$\rep{1}_{-2,0}$ & $H^{\bullet}(X,\mathcal{K}\otimes \mathcal{E}^*=L_1^{2})$ & $\tilde C_3$ \tabularnewline
			\midrule 
			$\rep{27}_{0,-2}$ & $H^{\bullet}(X,\mathcal{F}^*=L_2^{2})$  & $f_1$\tabularnewline
			\midrule
			$\rep{27}_{1,1}$ & $H^{\bullet}(X,\mathcal{K}^*=L_1^{-1}L_2^{-1})$  & $f_2$\tabularnewline
			\midrule
			$\rep{27}_{-1,1}$ & $H^{\bullet}(X,\mathcal{E}^*=L_1L_2^{-1})$  & $f_3$\tabularnewline
			\midrule
			$\rep{27}_{0,2}$ & $H^{\bullet}(X,\mathcal{F}=L_2^{-2})$  & $\tilde f_1$\tabularnewline
			\midrule
			$\rep{27}_{-1,-1}$ & $H^{\bullet}(X,\mathcal{K}=L_1L_2)$  & $\tilde f_2$\tabularnewline
			\midrule
			$\rep{27}_{1,-1}$ & $H^{\bullet}(X,\mathcal{E}=L_1^{-1}L_2)$  & $\tilde f_3$\tabularnewline
			\bottomrule
		\end{tabular}
		\par\end{centering}
		\caption{Low-energy representations of $E_6\times U(1)\times U(1)$ and their associated cohomologies. $L_1$ is a line bundle of the form $L_1=\mathcal{O}_X(m^1,m^2,m^3)$, while we write $L_2=\mathcal{O}_X(n^1,n^2,n^3)$ for $L_2$ . The entries in the third column correspond to either a chiral or an anti-chiral supermultiplet if the Euler characteristic of the associated line bundle is negative or positive respectively. Hence, the supermultiplets corresponding to a line bundle and its inverse bundle are conjugates of each other.
Note that the fields $C_i$ and $\tilde C_i$ are singlets under $E_6$. Deforming the bundle $V_{\rep 3}$ away from the decomposable locus is equivalent to turning on different combinations of VEVs for the scalar components of these supermultiplets in the effective theory.}
		\label{tab:4thTable}
\end{table}

\subsection*{Extension Bundle}

In this section, we study how to deform the Whitney sum bundle defined in eq.~\eqref{eq:defineBundle2} away from the decomposable locus to construct an irreducible $\SU3$ bundle. We will also discuss how this construction appears in the effective field theory.

Consider the pair of exact sequences
\begin{equation}
\begin{split}
\label{eq:sequences}
0&\rightarrow \mathcal{F}\rightarrow W\rightarrow \mathcal{E}\rightarrow  0\ ,\\
0&\rightarrow \mathcal{K}\rightarrow V^{\prime}_{\rep 3} \rightarrow W \rightarrow 0\ .
\end{split}
\end{equation}
These define an $SU(3)$ bundle $V^{\prime}_{\rep 3}$, since $c_1(V^{\prime}_{\rep 3})=0$. The first extension can be non-trivial -- that is, $W$ is not simply $\mathcal{F}\oplus\mathcal{E}$ -- if and only if the Ext group $\op{Ext}^1(\mathcal{E},\mathcal{F})=H^1(X, \mathcal{F}\otimes \mathcal{E}^*)$ is non-trivial. Similarly, the second extension can be non-trivial if and only if $\op{Ext}^1(W,\mathcal{K})=H^1(X,\mathcal{K}\otimes W^*)$ is non-trivial. It is relatively easy to see that a non-trivial extension class for the first extension sequence corresponds to turning on a VEV for the field $C_2$ in the effective theory. Indeed, from Table \ref{tab:4thTable} we see that $H^1(X, \mathcal{F}\otimes \mathcal{E}^*)$ counts the number of $C_2$ fields. It is straightforward, if not obvious, to see that the extension class for the second sequence is equivalent to turning on a VEV for the field $\tilde C_3$. To prove this, note that one can show that $\op{Ext}^1(W,\mathcal{K})=H^1(X,\mathcal{K}\otimes W^*)=H^1(X,\mathcal{K}\otimes \mathcal{E}^*)$~\cite{Anderson:2010ty}.\footnote{Specifically, in Section 3.3 of \cite{Anderson:2010ty} it is
shown that $H^1(X,\mathcal{K}\otimes W^*)=H^1(X,\mathcal{K}\otimes \mathcal{E}^*)+\kernel \delta$, where $\kernel \delta\in H^1(X,\mathcal{K}\otimes \mathcal{F}^*)$. Note that this branch is confined to the zero element of $H^1(X,\mathcal{K}\otimes \mathcal{F}^*)$.}  It follows from Table \ref{tab:4thTable} that $H^1(X, \mathcal{K}\otimes \mathcal{E}^*)$ counts the number of $\tilde C_3$ fields.
We learn that the pair of non-trivial extensions in \eqref{eq:sequences} corresponds to turning on VEVs for the fields $ C_2$ and $\tilde C_3$ in the effective theory,
\begin{equation}
\langle  C_2\rangle \neq 0\ , \qquad \langle \tilde C_3\rangle \neq 0 \ ,
\end{equation}
while the VEVs of the other charged matter fields are set to zero. Having allowed for these VEVs, one should check that the F-term and D-term constraints are satisfied so that the vacuum of the four-dimensional theory preserves supersymmetry. As shown in \cite{Anderson:2010ty}, one can turn on this combination of VEVs while still having the superpotential and its derivative vanish -- that is, the effective theory satisfies the F-flatness requirement. Furthermore, using the generic expression for a D-term given in \cite{Anderson:2010ty}, we see that fields $C_2$ and $\tilde C_3$ have the correct charges to cancel the FI terms associated with the two $U(1)$ bundles $\mathcal{F}$ and $\mathcal{K}$, as defined in Appendix \ref{app:FI}. Therefore,
the theory is also D-flat.

Note that there is another pair of extension sequences which correspond to the same non-zero elements of $H^1(X,\mathcal{F}\otimes \mathcal{E}^*)$ and $H^1(X,\mathcal{K}\otimes \mathcal{E}^*)$, namely
\begin{align}\label{eq:sequences_disc2}
	\begin{split}
		 0&\rightarrow \mathcal{K}\rightarrow W^\prime\rightarrow \mathcal{E}\rightarrow  0\ ,\\
			0&\rightarrow \mathcal{F}\rightarrow \mathcal{V}^{\prime}_{\rep 3} \rightarrow W^\prime \rightarrow 0\ ,
	\end{split}
	\begin{split}
		\op{Ext}^1(\mathcal{E},\mathcal{K})&=H^1(X,\mathcal{K}\otimes \mathcal{E}^*)\neq 0\ ,\\
		\op{Ext}^1(W^\prime,\mathcal{F})&=H^1(X,\mathcal{F}\otimes \mathcal{E}^*) \neq 0\ .
	\end{split}
\end{align}
This seemingly gives us a different $SU(3)$ bundle $\mathcal{V}^{\prime}_{\rep 3}$ for the same set of VEVs. However, as shown in \cite{Anderson:2010ty}, this new bundle is actually isomorphic to the above, that is
\begin{equation}
\mathcal{V}^{\prime}_{\rep 3}\simeq V^{\prime}_{\rep 3}\ .
\label{cold1}
\end{equation}
From an effective theory perspective, both pairs of sequences correspond to turning on VEVs for the low-energy fields $\tilde{C}_2$ and $C_3$. Roughly speaking, the choice between \eqref{eq:sequences} or \eqref{eq:sequences_disc2} corresponds to turning on VEVs for $C_2$ first and then $\tilde{C}_3$ or vice versa, respectively.

As we will discuss in Section \ref{sec:different_branches} there are actually six inequivalent \emph{branches} along which one can, in principle, deform the $SU(3)$ bundle away from the decomposable locus. From the low-energy perspective, deforming the bundle along one branch or another comes from turning on different pairs of VEVs of the $E_6$ singlet fields $C_i$ and $\tilde C_i$ which are both F-flat and D-flat. The extensions in \eqref{eq:sequences} or \eqref{eq:sequences_disc2} give one possible branch. 
We will study the full branch structure for a hidden sector built from two line bundles in detail in Sections \ref{sec:different_branches} and \ref{sec:full_scan}, but for the moment let us continue focusing on the $\langle  C_2\rangle,\langle \tilde C_3\rangle \neq 0$ branch. We have seen that a non-trivial extension is possible if and only if the cohomology groups $H^1(X,\mathcal{F}\otimes \mathcal{E}^*)$ and $H^1(X,\mathcal{K}\otimes \mathcal{E}^*)$ are non-empty, that is
\begin{equation}
h^1(X,\mathcal{F}\otimes \mathcal{E}^*)>0\ , \qquad h^1(X,\mathcal{K}\otimes \mathcal{E}^*)>0 \ .
\end{equation}
It is useful to observe that the case where either of the line bundles $\mathcal{F}\otimes \mathcal{E}^*$ or $\mathcal{K}\otimes \mathcal{E}^*$ is ample is eliminated from the start, since ample line bundles have vanishing first cohomology groups on a Calabi--Yau manifold. Hence, one must choose the line bundles $L_{1}$, $L_{2}$ so that $\mathcal{F}\otimes \mathcal{E}^*$ and $\mathcal{K}\otimes \mathcal{E}^*$ are {\it not} ample. That being said, computing the cohomology groups $H^1(X,\mathcal{F}\otimes \mathcal{E}^*)$ and $H^1(X,\mathcal{K}\otimes \mathcal{E}^*)$ directly on our Schoen manifold $X$ is a difficult task, which far exceeds the scope of the present paper. This will be discussed in future publications.

\subsection*{Bundle Stability}

The conditions above ensure that an extension is possible, which corresponds to the existence of non-zero VEVs of certain charged matter fields in the four-dimensional effective theory. The existence of the deformed $SU(3)$ bundle is not sufficient to ensure supersymmetry however. One must also require that the new bundle admits a connection that solves the HYM equation; that is, the bundle must be slope-stable. Therefore, the next question one must ask, assuming the extension exists, is whether the resulting bundle is slope-stable. As we have emphasized, checking stability of a bundle is generally a difficult calculation which, at the moment, cannot be done algorithmically on our particular Schoen threefold. Instead, we will focus on some necessary conditions; specifically that some obvious subbundles have negative slope and that the Bogomolov inequality is satisfied. This last requirement is significant  since often the Bogomolov inequality is the only obstruction to finding a slope-stable bundle~\cite{Braun:2006ae}. 

From the first sequence in \eqref{eq:sequences}, we learn that there is an embedding
\begin{equation}
\mathcal{K}\hookrightarrow V^\prime_{\rep 3}\ .
\end{equation}
That is, the line bundle $\mathcal{K}$ injects into $V_{\rep 3}^\prime$. Since $V_{\rep 3}^\prime$ has vanishing slope, the first necessary condition for stability is that $\mathcal{K}$ has negative tree-level slope:
\begin{equation}
\label{eq:slopecond1}
\mu(\mathcal{K})=\mu(L_1L_2)<0\quad \Rightarrow \quad d_{ijk}(m^i+n^j)a^ja^k<0\ ,
\end{equation}
Furthermore, it would appear from \eqref{eq:sequences_disc2} and \eqref{cold1}, and is proven in Appendix \ref{app:subbundles}, that the bundle $\mathcal{F}$ also injects into $ V^\prime_{\rep 3}$,
\begin{equation}
\mathcal{F}\hookrightarrow V^\prime_{\rep 3}\ ,
\end{equation}
and so the slope of $\mu(\mathcal{F})$ must also be negative:
\begin{equation}
\label{eq:slopecond2}
\mu(\mathcal{F})=\mu(L_2^{-2})<0\quad \Rightarrow \quad -2d_{ijk}n^ja^ja^k<0\ .
\end{equation}

Given a choice of $L_1$ and $L_2$ such that \eqref{eq:slopecond1} and \eqref{eq:slopecond2} are satisfied, the final necessary condition that we impose is that $V_{\rep 3}^{\prime}$ satisfies the Bogomolov inequality
\begin{equation}
\int c_{2}(V_{\rep 3}^{\prime})\wedge\omega\geq0\ .
\label{train1}
\end{equation}
A bundle is stable only if it satisfies this highly non-trivial constraint. Thankfully, since $c_{2}$ is topological, this can be computed from the data of the Whitney sum bundle $V_{\rep 3}=\mathcal{F}\oplus\mathcal{K}\oplus\mathcal{E}$. The total Chern class is
\begin{equation}
\begin{split}
c(V_{\rep 3}^{\prime})\equiv c(V_{\rep 3}) & =c( L_2^{-2}\oplus L_1L_2\oplus L_1^{-1}L_2)\\
 &=1-3c_1^2(L_2)-c_1^2(L_1)+\dots
 \end{split}
 \end{equation}
from which we see that $c_2(V_{\rep 3}^\prime)=3c_1^2(L_2)-c_1^2(L_2)$. It follows that condition \eqref{train1} becomes
\begin{equation}
\label{eq:bogomolov_2lines}
d_{ijk}m^im^ja^k+3 d_{ijk}n^in^ja^k\leq 0\ .
\end{equation}
We emphasize that since the K\"ahler form $\omega$ appears in the calculation of the two slopes and the Bogomolov inequality, all three of these conditions depend on where one is in the K\"ahler cone, which is also restricted by the physical constraints discussed in Section \ref{Physical Constraints2}.

\subsection{Line Bundles Scan}\label{sec:first_survey}

We are now in a position to look for an appropriate hidden sector bundle, which has a non-trivial extension $V_{\rep 3}^\prime$ that might be stable away from the decomposable locus. In addition to the physical constraints \eqref{51}--\eqref{eq:linearizedCons}, we impose the constraints derived so far in this section. Specifically, we want to find two line bundles $L_1=\mathcal{O}_X(m^1,m^2,m^3)$ and $L_2=\mathcal{O}_X(n^1,n^2,n^3)$
such that

\begin{enumerate}
\item Both $L_1$ and $L_2$ are equivariant:
\begin{equation}
(m_1+m_2)\op{mod}3=0\ , \quad (n_1+n_2)\op{mod}3=0\ .
\end{equation}

\item The five-brane class is effective:
\begin{equation}
W_i=\left(\tfrac{4}{3},\tfrac{7}{3},-4  \right)_i+d_{ijk}m^jm^k+3d_{ijk}n^jn^k\geq0\ .
\end{equation}

\item The cohomology groups $H^1(X, \mathcal{F}\otimes \mathcal{E}^*)=H^1(X, L_1L_2^{-3})$ and $H^1(X, \mathcal{K}\otimes \mathcal{E}^*)=H^1(X, L_1^2)$ are non-zero. In this scan we will simply impose the necessary condition that $\mathcal{F}\otimes \mathcal{E}^*=L_1L_2^{-3}$ and $ \mathcal{K}\otimes \mathcal{E}^*=L_1^2$ are not ample.

\item The extended bundle $V_{\rep 3}^\prime$ is stable. The necessary conditions that we impose are:

(i) The slopes of the line bundles $\mathcal{F}$ and $\mathcal{K}$ are negative:
\begin{equation}
\label{eq:const_2slopes}
 d_{ijk}(m^i+n^i)a^j a^k<0 \ , \qquad -2d_{ijk}n^i a^j a^k<0\ .
\end{equation}

(ii) The extension bundle $V_{\rep 3}^\prime$ satisfies the Bogomolov inequality
\begin{equation}
\label{eq:bogolomov_cons2}
d_{ijk}m^im^ja^k+3 d_{ijk}n^in^ja^k\leq 0\ .
\end{equation}

\end{enumerate}

\begin{figure}[t]
   \centering
     \begin{subfigure}[b]{0.46\textwidth}
\includegraphics[width=1.0\textwidth]{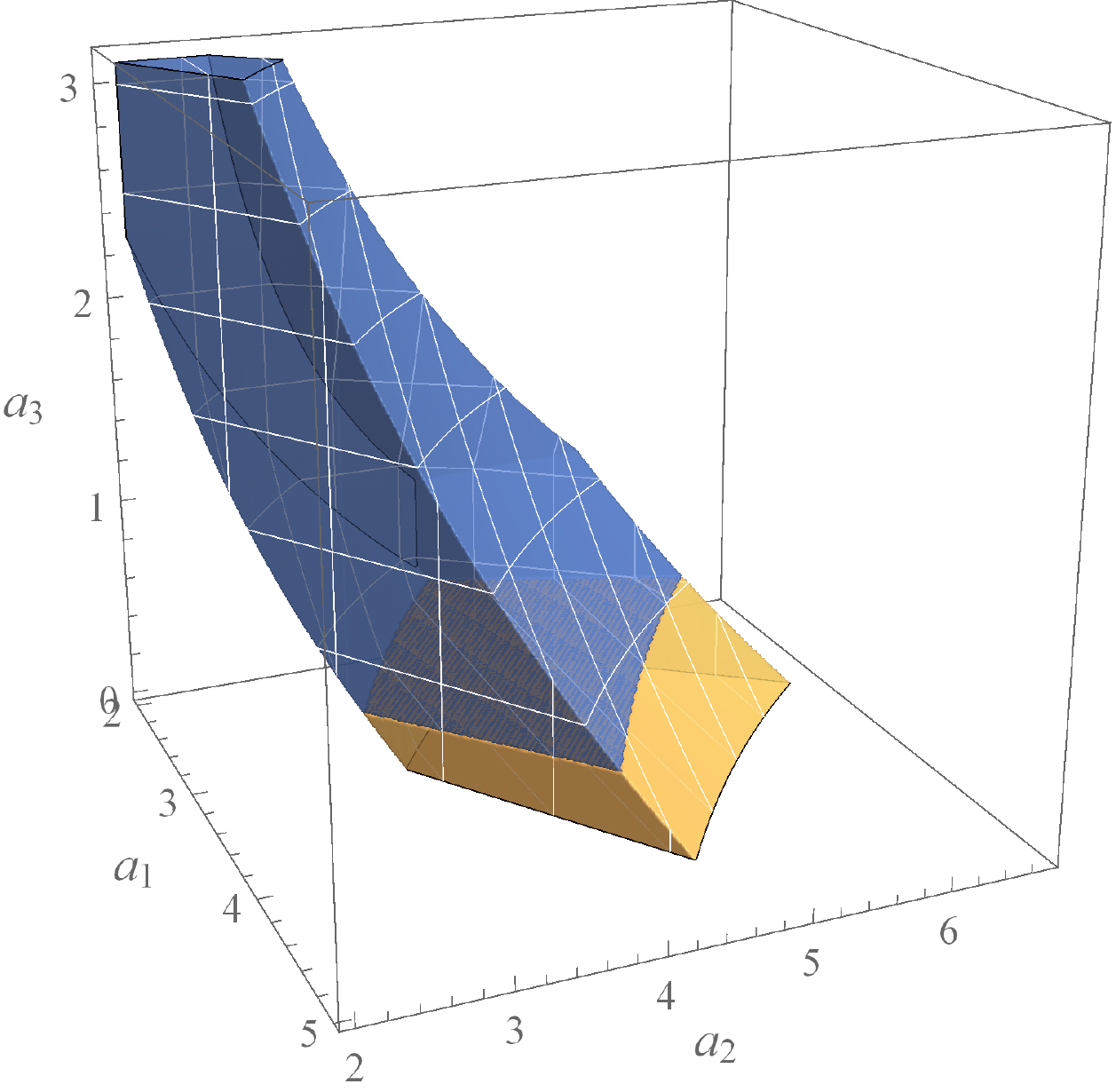}
\end{subfigure}
\caption{ The colored region is the subspace of K\"ahler moduli space which, in addition to satisfying conditions \eqref{51}-\eqref{jack1} outlined in Section \ref{Physical Constraints}, also restricts the strong coupling parameter to the smaller values $\epsilon_S^{\text{eff}} \lesssim 0.2$. Hence, any point in this region also satisfies the linearization constraint \eqref{eq:linearizedCons}. This subspace does not depend on the
hidden sector bundle.
The ``orange'' regime is a subspace of this colored region, in which the non-trivial $SU(3)$ bundle $V_{\rep 3}^\prime$, defined as an extension of the two line bundles $L_1=\mathcal{O}_X(-5, -1 ,1)$ and $L_2=\mathcal{O}_X(2, 1, -1)$,
can be slope-stable. Note that one can show that the locus where the {\it decomposable} Whitney sum bundle $V_{\rep 3}$ is poly-stable lies outside the colored region shown in Figure \ref{fig:ModuliSolution_orange} and, hence, does not satisfy all of the universal constraints.
}
\label{fig:moduliSolution}
\end{figure}

We performed a systematic scan over all possible pairs of line bundles, with $|m^{i}|\leq 15$, $|n^{i}|\leq 15$ for $i=1,2,3$. The values of $(a^1,a^2,a^3)$ we sample sit inside the ``blue'' subspace of the K\"ahler cone shown in Figure \ref{fig:ModuliSolution_orange}. 
We find a number of pairs of line bundles which lead to a solution satisfying all of these constraints. These include the line bundles
\begin{equation}
L_1=\mathcal{O}_X(-5, -1, 1)\ , \qquad  L_2=\mathcal{O}_X(2,1, -1).
\label{door1}
\end{equation}
Before giving the other examples, let us analyze this case in more detail. For this pair of line bundles, we find that the class of the five-brane is
\begin{equation}
W_i=(2,  0, 18)_i\ ,
\end{equation}
which is indeed effective ($W_i\geq 0$).
Furthermore, the bundles $L_1L_2^{-3}=\mathcal{O}_X(-11,-4,4)$ and $L_1^2=\mathcal{O}_X(-10,-2,2)$, associated with the $\rep 1_{-1,3}$ and the $\rep 1_{-2,0}$ representations respectively, are clearly non-ample. However, we do not know if the extensions in \eqref{eq:sequences} exist without computing the cohomologies
\begin{align}
\label{eq:cohomology}
&H^1(X, \mathcal{F}\otimes \mathcal{E}^*)=H^1(X, L_1L_2^{-3})=H^1(X, \mathcal{O}_X(-11,-4,4)).\\
&H^1(X, \mathcal{K}\otimes \mathcal{E}^*)\equiv H^1(X, L_1^{2})=H^1(X, \mathcal{O}_X(-10,-2,2))\ .
\end{align}
Assuming these cohomologies are non-empty, it is simple to check that the necessary conditions for stability are also satisfied.

In order to display the region of stability of the extension $V_{\rep 3}^\prime$ of the line bundles \eqref{door1}, we sample values $(a^1,a^2,a^3)$ inside the ``blue'' region in Figure \ref{fig:ModuliSolution_orange} and check the stability criteria we gave in equations \eqref{eq:const_2slopes} and \eqref{eq:bogolomov_cons2}. Importantly, we choose to restrict our discussion to the ``blue'' region in Figure \ref{fig:ModuliSolution_orange} since in this subspace, by definition, the strong coupling parameter satisfies $\epsilon_S^{\text{eff}} \lesssim 0.2$ and, hence, is relatively small. As discussed in Section 2, this implies that the linearization constraint \eqref{eq:linearizedCons} is satisfied. The subset of the blue region in Figure \ref{fig:ModuliSolution_orange} where the extension bundle can be slope-stable is displayed as the ``orange'' subspace in Figure \ref{fig:moduliSolution}.  Having found  stable bundles $V_{\rep 3}^\prime$ inside this ``orange''
 \begin{figure}[t]
   \centering
     \begin{subfigure}[b]{0.49\textwidth}
\includegraphics[width=1.0\textwidth]{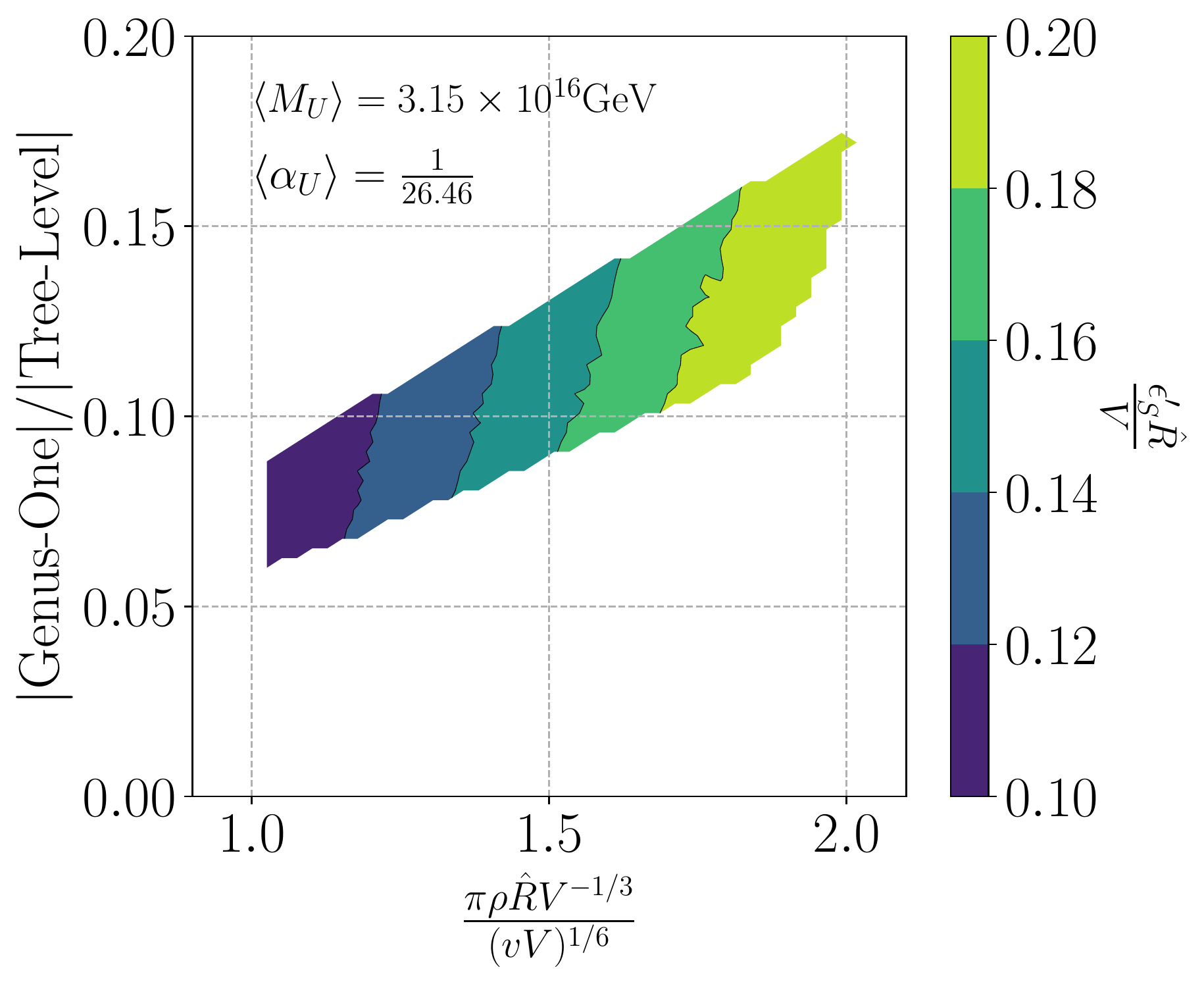}
\caption{}
\end{subfigure}
     \begin{subfigure}[b]{0.49\textwidth}
\includegraphics[width=1.0\textwidth]{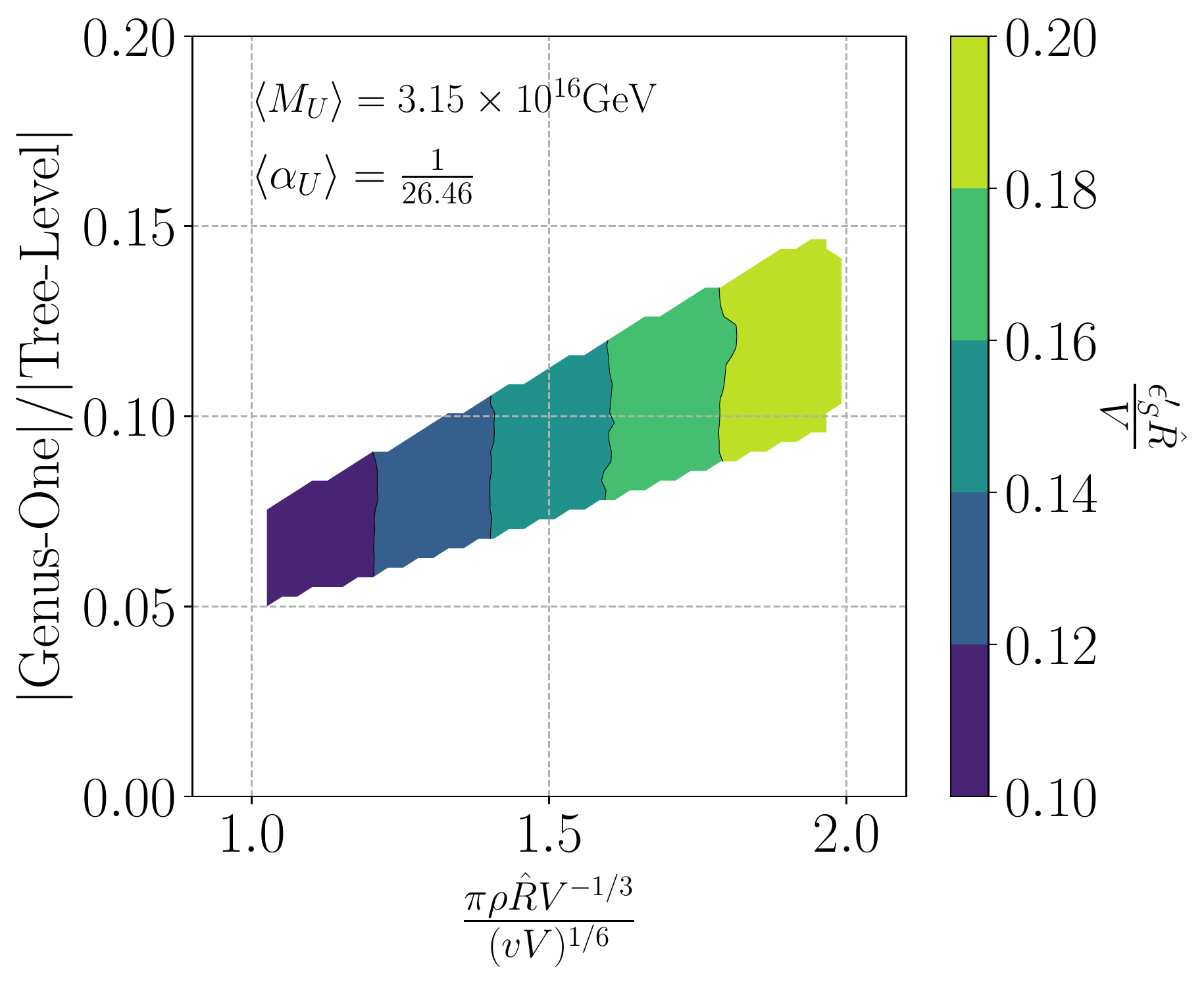}
\caption{}
\end{subfigure}
\caption{In this Figure, for the line bundles $L_1=\mathcal{O}_X(-5, -1, 1)$ and $L_2=\mathcal{O}_X(2,1, -1)$, we plot both the ratio of the genus-one correction to
the tree-level value of the FI-term associated with the bundles $\mathcal{F}$ in (a) and $\mathcal{K}$ in (b), against the ratio of the fifth-dimensional length $\pi \rho \hat R V^{-1/3}$
to the average Calabi–Yau radius $(vV)^{-1/6}$ 
(displayed
on the horizontal axis). We also show the values of the effective strong
coupling parameter $\epsilon_S^{\text{eff}}=\epsilon_S^\prime\hat R/V$ using the colored shading. Points in the plot are sampled from inside the ``orange'' region of Figure \ref{fig:moduliSolution}. Across the solution space, the ratios from \eqref{eq:ratio_genus_F} and \eqref{eq:ratio_genus_K} are less that $0.2$. Furthermore, in both cases, the effective strong coupling parameter $\epsilon_S^{\text{eff}}$ is no larger than $0.2$, suggesting that the linear approximation is accurate.}
\label{fig:2D_plot}
\end{figure} 
\noindent region allows us to move away from the decomposable locus and, importantly, away from the region of K\"ahler moduli space in which the theory is very strongly coupled.

In addition to the strong coupling parameter satisfying $\epsilon_S^{\text{eff}} \lesssim 0.2$, there are two other criteria that specify whether or not the strong coupling corrections are relatively small. First, let us analyze the ratio 
of the genus one corrections to the tree level value of the Fayet-Iliopoulos term associated with each of the two line {\it generic} bundles $\mathcal{F}$ and $\mathcal{K}$.
As shown in Appendix \ref{app:FI}, the FI-terms associated with $\mathcal{F}$ and $\mathcal{K}$ are given by
\begin{equation}
\begin{split}
\label{eq:FI_K}
FI_{\mathcal{F}}&=\frac{1}{2}\frac{\epsilon_S\epsilon_R^2}{\kappa_4^2}\frac{1}{\hat R V^{2/3}}\left[-2d_{ijk}n^ia^ja^k+\frac{2\epsilon_S^\prime \hat R}{V^{1/3}}n^i \Big(   \big( \tfrac{2}{3},-\tfrac{1}{3},4 \big)\big|_i  +\left(1-(1+\tfrac{\lambda}{2})^2\right) W_i    \Big)\right]\ ,\\
FI_{\mathcal{K}}&=\frac{1}{2}\frac{\epsilon_S\epsilon_R^2}{\kappa_4^2}\frac{1}{\hat R V^{2/3}}\bigg[d_{ijk}(m^i+n^i)a^ja^k\\
&\eqspace\qquad\qquad-\frac{\epsilon_S^\prime \hat R}{V^{1/3}}(m^i+n^i) \Big(   \big( \tfrac{2}{3},-\tfrac{1}{3},4 \big)\big|_i  +\left(1-(1+\tfrac{\lambda}{2})^2\right) W_i    \Big)\bigg]\ .\\
\end{split}
\end{equation}
Making use of ``unity'' gauge where  one sets $\epsilon_S^\prime \hat R/V^{1/3}=1$, we find that this ratio is given by
\begin{align}
\label{eq:ratio_genus_F}
\left|\frac{\text{Genus-One}}{\text{Tree-Level}}\right|_{\mathcal{F}}=\frac{\left|-2n^i \Big(   \big( \tfrac{2}{3},-\tfrac{1}{3},4 \big)\big|_i  +\left(1-(1+\tfrac{\lambda}{2})^2\right) W_i    \Big)\right|}{|2d_{ijk}n^i a^j a^k|}\ ,
\end{align}

\begin{align}
\label{eq:ratio_genus_K}
\left|\frac{\text{Genus-One}}{\text{Tree-Level}}\right|_{\mathcal{K}}=\frac{\left|(m^i+n^i) \Big(   \big( \tfrac{2}{3},-\tfrac{1}{3},4 \big)\big|_i  +\left(1-(1+\tfrac{\lambda}{2})^2\right) W_i    \Big)\right|}{|d_{ijk}(m^i+n^i)a^j a^k|} 
\end{align}
for $\mathcal{F}$ and $\mathcal{K}$ respectively.
A second criterion that the strong coupling corrections are relatively small is that ratio between the orbifold length and the average Calabi--Yau radius, given by
\begin{equation}
 \frac{\pi \rho \hat R V^{-1/3}}{(vV)^{-1/6}} 
\label{bird1}
\end{equation}
where $V=\frac{1}{6} d_{ijk}a^{i}a^{j}a^{k}$, should not exceed a value of $\sim 2.$  In \cite{Ashmore:2020ocb} it was shown that increasing the orbifold length, so that the hidden and observable sectors are more separated, also increased the effective expansion parameter, causing the linear approximation to become less accurate. For an effective expansion parameter of order 1, the orbifold length is roughly 12 times larger than the Calabi--Yau scale. 

Returning to our explicit solution with line bundles $L_1=\mathcal{O}_X(-5, -1, 1)$ and $L_2=\mathcal{O}_X(2,1, -1)$ given in \eqref{door1}, we now plot \eqref{eq:ratio_genus_F}, \eqref{eq:ratio_genus_K} and \eqref{bird1}, as well as $\epsilon_S^{\text{eff}}=\epsilon_S^\prime\hat R/V$, over all points in the ``orange'' region of Figure 2. 
We find that across the ``orange'' solution space, in the region where the ratio between the orbifold length and the average Calabi--Yau radius is less than 2, the genus-one corrections to the FI terms of $\mathcal{F}$ and $\mathcal{K}$ in eq.~\eqref{eq:FI_K} are less than $0.2$ times the tree-level values. Finally, the effective strong coupling parameter $\epsilon_S^{\text{eff}}$ is no larger than $0.2$, as expected since we are within the ``blue'' region of Figure 1. We conclude that for K\"ahler moduli in the ``orange'' regime of Figure 2, all three criteria for relatively small $\kappa_{11}^{4/3}$ corrections are satisfied. 
It is of interest to compare these results to the single line bundle case studied in \cite{Ashmore:2020ocb}, where cancelling the tree-level FI term against the genus-one correction was necessary to obtain a poly-stable hidden sector bundle. In that context, ${|\text{Genus-One}|}/{|\text{Tree-Level}|}|\approx 1$, while the effective expansion parameter $\epsilon_S^{\text{eff}}=\epsilon_S^\prime\hat R/V$ was also of order 1. Hence, in that case, the linear approximation was less trustworthy. That is, the results shown in Figure \ref{fig:2D_plot} are well within the range of applicability of the linear approximation, which is a significant improvement over \cite{Ashmore:2020ocb}.

For a particular branch of the hidden sector bundle construction, we have found at least one example of two line bundles $(L_1,L_2)$ which can potentially solve all of the geometrical, phenomenological and dimensional reduction constraints we have outlined in this paper. The problem of finding viable hidden sectors has thus been reduced to the calculation of the $H^1(X,  L_1L_2^{-3})$ and $H^1(X, L_1^{2})$ cohomologies (modulo whether the necessary conditions for stability that we have imposed do indeed lead to stable bundles). The two bundles $L_1=\mathcal{O}_X(-5, -1, 1)$ and $L_2=\mathcal{O}_X(2,1, -1)$ that we have discussed in detail in this section are not unique -- there are actually a large number of pairs of line bundles that satisfy the same constraints. We show a subset of them in Table 
\ref{tab:solutions_branch1}, restricted to the range $|m^{i}|\leq 15$, $|n^{i}|\leq 15$, with $i=1,2,3$.

Working with two line bundles has introduced more degrees of freedom into our system and, implicitly, more flexibility in finding solutions than was present in the single line bundle case given in \cite{Ashmore:2020ocb}. With this extra freedom, we have potentially succeeded in building a non-trivial $SU(3)$ bundle $V_{\rep 3}^\prime$, which is stable within the physically viable region of the Kähler cone.

\begin{table}[t]
	\noindent \begin{centering}
		\begin{tabular}{cc}
			\toprule 
			$L_1$ & $L_2$     \tabularnewline
						\midrule
			\midrule 
			$\mathcal{O}_X(-11,-10,4)$ & $\mathcal{O}_X(-2,14,-2)$       \tabularnewline
			\midrule 
			 $\mathcal{O}_X(-8,-7,3)$ & $\mathcal{O}_X(-1,13,-2)$    \tabularnewline
			 \midrule
			  $\mathcal{O}_X(-5,-4,2)$ & $\mathcal{O}_X(2,-2,0)$       \tabularnewline
			 \midrule
			 $\mathcal{O}_X(-5,-1,1)$ & $\mathcal{O}_X(2,1,-1)$       \tabularnewline
			 \midrule
		\end{tabular}
		\par\end{centering}
		\caption{A subset of pairs of line bundles that lead to solutions of the full set of the constraints on the branch $\langle C_2\rangle, \langle \tilde C_3\rangle\neq 0$. The solution space which corresponds to $L_1=\mathcal{O}_X(-5, -1 ,1)$ and $L_2=\mathcal{O}_X(2, 1, -1)$ was shown in Figure \ref{fig:moduliSolution}. We have restricted our scan to line bundles $L_1=\mathcal{O}_X(m^1,m^2,m^3)$ and $L_2=\mathcal{O}_X(n^1,n^2,n^3)$ for which $|m^{i}|\leq 15$, $|n^{i}|\leq 15$.}
		\label{tab:solutions_branch1}
\end{table}

\section{Different Extension Branches}\label{sec:more_branches}

\begin{table}[t]
	\noindent \begin{centering}
		\begin{tabular}{cccc}
			\toprule 
			Branch & Field VEVs & Sequences & Ext\tabularnewline
						\midrule
			\midrule 
			1& $\langle C_2\rangle,\langle \tilde C_3 \rangle \neq 0$&
			 $\begin{matrix} 0\rightarrow \mathcal{F} \rightarrow W\rightarrow \mathcal{E}\rightarrow 0\\ 0\rightarrow \mathcal{K} \rightarrow V_{\rep 3}^\prime \rightarrow W\rightarrow 0 \end{matrix} $  &  
			 $\begin{matrix}H^1(X,\mathcal{F}\otimes \mathcal{E}^*)\\H^1(X,\mathcal{K}\otimes \mathcal{E}^*)\end{matrix}$\tabularnewline
			 \midrule
			 	2& $\langle C_3\rangle,\langle \tilde C_2 \rangle \neq 0$&
			 $\begin{matrix} 0\rightarrow \mathcal{E} \rightarrow W^{(2)}\rightarrow \mathcal{K}\rightarrow 0\\ 0\rightarrow W^{(2)} \rightarrow {V_{\rep 3}^\prime}^{(2)} \rightarrow \mathcal{F}\rightarrow 0 \end{matrix} $  &  
			 $\begin{matrix}H^1(X,\mathcal{E}\otimes \mathcal{K}^*)\\H^1(X,\mathcal{E}\otimes \mathcal{F}^*)\end{matrix}$\tabularnewline
			 \midrule
			 	3& $\langle C_2\rangle,\langle  C_1 \rangle \neq 0$&
			 $\begin{matrix} 0\rightarrow \mathcal{F} \rightarrow W^{(3)}\rightarrow \mathcal{E}\rightarrow 0\\ 0\rightarrow W^{(3)} \rightarrow {V_{\rep 3}^\prime}^{(3)} \rightarrow \mathcal{K}\rightarrow 0 \end{matrix} $  &  
			 $\begin{matrix}H^1(X,\mathcal{F}\otimes \mathcal{E}^*)\\H^1(X,\mathcal{F}\otimes \mathcal{K}^*)\end{matrix}$\tabularnewline
			 \midrule
			 	4& $\langle \tilde C_1\rangle,\langle \tilde C_2 \rangle \neq 0$&
			 $\begin{matrix} 0\rightarrow \mathcal{K} \rightarrow W^{(4)}\rightarrow \mathcal{F}\rightarrow 0\\ 0\rightarrow \mathcal{E} \rightarrow {V_{\rep 3}^\prime}^{(4)} \rightarrow W^{(4)}\rightarrow 0 \end{matrix} $  &  
			 $\begin{matrix}H^1(X,\mathcal{K}\otimes \mathcal{F}^*)\\H^1(X,\mathcal{E}\otimes \mathcal{F}^*)\end{matrix}$\tabularnewline
			 \midrule
			 	5& $\langle C_3\rangle,\langle C_1 \rangle \neq 0$&
			 $\begin{matrix} 0\rightarrow \mathcal{E} \rightarrow W^{(5)}\rightarrow \mathcal{K}\rightarrow 0\\ 0\rightarrow \mathcal{F} \rightarrow {V_{\rep 3}^\prime}^{(5)} \rightarrow W^{(5)}\rightarrow 0 \end{matrix} $  &  
			 $\begin{matrix}H^1(X,\mathcal{E}\otimes \mathcal{K}^*)\\H^1(X,\mathcal{F}\otimes \mathcal{K}^*)\end{matrix}$\tabularnewline
			 \midrule
			 	6& $\langle \tilde C_1\rangle,\langle \tilde C_3 \rangle \neq 0$&
			 $\begin{matrix} 0\rightarrow \mathcal{K} \rightarrow W^{(6)}\rightarrow \mathcal{F}\rightarrow 0\\ 0\rightarrow W^{(6)} \rightarrow {V_{\rep 3}^\prime}^{(6)} \rightarrow \mathcal{E} \rightarrow 0 \end{matrix} $  &  
			 $\begin{matrix}H^1(X,\mathcal{K}\otimes \mathcal{F}^*)\\H^1(X,\mathcal{K}\otimes \mathcal{E}^*)\end{matrix}$\tabularnewline
			\bottomrule
		\end{tabular}
		\par\end{centering}
		\caption{The six extension branches of the split Whitney sum bundle $V_{\rep 3}=\mathcal{F}\oplus \mathcal{K}\oplus
		\mathcal{E}$. For each branch, there is also a second pair of sequences which corresponds to switching the order of the extensions. The resulting bundles can be shown to be isomorphic~\cite{Anderson:2010ty} and, hence, we do not display them.}
		\label{tab:branches}
\end{table}

In the example in the previous section, we showed it might be possible to extend the decomposable bundle $V_{\rep 3}$, defined in \eqref{eq:defineBundle2}, to a non-trivial stable $SU(3)$ bundle $V_{\rep 3}^\prime$ via the extension sequences defined in \eqref{eq:sequences}. As we discussed, this extension is equivalent to turning on VEVs for the fields $C_2$ and $\tilde C_3$ in the four-dimensional effective theory. We now ask if we can still solve the system of vacuum constraints if we chose a different extension sequence or, equivalently, if we chose to turn on VEVs for different combinations of $C$ and $\tilde C$ fields. 

First of all, not all combinations of pairs of VEVs are allowed. The F-flatness conditions, coming from the vanishing of the superpotential and its first derivative, reduce the fifteen combinations of pairs of VEVs to six~\cite{Anderson:2010ty}. In Table \ref{tab:branches} we give the allowed VEVs for each branch and the corresponding extension sequences. The branch we studied in the previous section corresponds to branch one in this table. Let us now analyse the conditions for a non-trivial extension and stability for the remaining branches. 

\subsection{Choosing a Different Branch}\label{sec:different_branches}

Consider first turning on VEVs for the fields $ C_1$ and $C_2$
\begin{equation}
\langle  C_1\rangle \neq 0\ , \qquad \langle C_2\rangle \neq 0\ ,
\end{equation}
while all other VEVs are set to zero. This combination of VEVs corresponds to the third branch in Table \ref{tab:branches}. The $U(1)\times U(1)$ bundle $V_{\rep 3}$ is deformed to an irreducible $SU(3)$ bundle ${V_{\rep 3}^\prime}^{(3)}$--where the superscript (3) indicates that one is working in the third branch of Table 3--via two extension sequences,
\begin{equation}
\begin{split}
\label{eq:sequences2}
\langle C_2\rangle\neq 0\colon\qquad  &0\rightarrow \mathcal{F}\rightarrow W^{(3)}\rightarrow \mathcal{E}\rightarrow  0\ ,\\
\langle C_1\rangle \neq 0\colon \qquad &0\rightarrow W^{(3)}\rightarrow {V^{\prime}}^{(3)}_{\rep 3} \rightarrow \mathcal{K} \rightarrow 0\ .
\end{split}
\end{equation}
In this case, a non-trivial extension ${V^{\prime}}^{(3)}_{\rep 3}$ exists if the Ext groups $\op{Ext}^1(\mathcal{E},\mathcal{F})=H^1(X, \mathcal{F}\otimes \mathcal{E}^*)$ and $\op{Ext}^1(\mathcal{K},\mathcal{F})=H^1(X, \mathcal{F}\otimes \mathcal{K}^*)$ are non-trivial. As for branch one, there is an alternative set of sequences which define an isomorphic bundle,
\begin{equation}
\begin{split}
\label{eq:sequences_disc4}
\langle C_1\rangle \neq 0\colon \qquad &0\rightarrow \mathcal{F}\rightarrow {W^\prime}^{(3)}\rightarrow \mathcal{K}\rightarrow  0\ , \\
\langle C_2\rangle\neq 0\colon\qquad  &0\rightarrow W^{(3)\prime}\rightarrow {\mathcal{V}^{\prime}}^{(3)}_{\rep 3} \rightarrow \mathcal{E} \rightarrow 0\ .
\end{split}
\end{equation}
Again, the cohomology groups $H^1(X,\mathcal{F}\otimes \mathcal{E}^*)$ and $H^1(X,\mathcal{F}\otimes \mathcal{K}^*)$ must be non-vanishing.
The final $SU(3)$ bundles are isomorphic, ${V^{\prime}}^{(3)}_{\rep 3} \sim {\mathcal{V}^{\prime}}^{(3)}_{\rep 3}$.

It follows from Table 1 that the fields $C_1$ and $C_2$  correspond to 
\begin{align}
\mathcal{F}\otimes \mathcal{K}^*= L_1^{-1}L_2^{-3}&=\mathcal{O}_X(-m_1-3n_1,-m_2-3n_2,-m_3-3n_3)\ ,\\
\mathcal{F}\otimes \mathcal{E}^*= L_1L_2^{-3}&=\mathcal{O}_X(m_1-3n_1,m_2-3n_2,m_3-3n_3)
\end{align}
respectively. Hence, as discussed above, in order for the cohomology groups  $H^1(X,\mathcal{F}\otimes \mathcal{E}^*)$ and $H^1(X,\mathcal{F}\otimes \mathcal{K}^*)$ to be non-vanishing, one must impose the constraint that 
the line bundles $L_1L_2^{-3}$ and $ L_1^{-1}L_2^{-3}$ be non-ample.

Assuming a non-trivial extension exists, we also want ${V_{\rep 3}^\prime}^{(3)}$ to be slope-stable.
Looking at the sequences in eq.~\eqref{eq:sequences2}, we see that the line bundle $\mathcal{F}$ injects into $W^{(3)}$, which itself injects into ${V_{\rep 3}^\prime}^{(3)}$. Hence $\mathcal{F}$ is a subbundle of ${V_{\rep 3}^\prime}^{(3)}$. Since ${V_{\rep 3}^\prime}^{(3)}$ has vanishing slope, it can be slope-stable only if $\mathcal{F}$ has negative slope at tree-level:
\begin{align}
\label{eq:slope_3rd}
\mu(\mathcal{F})=\mu(L_2^{-1})<0\quad \Rightarrow \quad -2d_{ijk}n^ia^ja^k<0\ .
\end{align}
We also obtain a condition on the slope of $\mathcal{K}$ from the same sequences in eq.~\eqref{eq:sequences2}. From the second line we learn that
\begin{equation}
c_1(W^{(3)})+c_1(\mathcal{K})=c_1({V_{\rep 3}^\prime}^{(3)})=0\ .
\label{bird2}
\end{equation} 
Recall that the slope of a bundle $\mathcal{L}$ is given by 
\begin{equation}
\label{eq:slope_form}
\mu(\mathcal{L})=\frac{1}{\rank\mathcal{L}}\int_X \omega\wedge \omega \wedge c_1(\mathcal{L})\ ,
\end{equation}
where $\rank \mathcal{L}$ is the rank of the bundle $\mathcal L$. Noting that $\rank \mathcal{K}=1$, $\rank{W}^{(3)}=2$ and $\rank V_{\rep 3}^{\prime (3)}=3$, it follows from \eqref{bird2} that
\begin{equation}
2\mu(W^{(3)})+\mu(\mathcal{K})=3\mu({V_{\rep 3}^\prime}^{(3)})=0\ .
\end{equation} 
Since $W^{(3)}$ is a subbundle of ${V_{\rep 3}^\prime}^{(3)}$, it must have a negative slope in order for $V_{\rep 3}^\prime$ to be slope-stable. Therefore, one must take the slope of $\mathcal{K}$ to be positive
\begin{align}
\label{eq:slopeK_3rd}
\mu(\mathcal{K})=\mu(L_1L_2)>0\quad \Rightarrow \quad 2d_{ijk}(m^i+n^i)a^ja^k>0\ .
\end{align}
From the alternative sequences in \eqref{eq:sequences_disc4}, we learn that 
\begin{equation}
2\mu(W^{(3)\prime})+\mu(\mathcal{E})=3\mu({\mathcal{V}_{\rep 3}^\prime}^{(3)})=0\ .
\end{equation} 
Again, since $W^{(3)\prime}$ is a subbundle of ${\mathcal{V}_{\rep 3}^\prime}^{(3)}$, its slope must be negative and, therefore, $\mathcal{E}$ must have positive slope. That is
\begin{align}
\label{eq:slopeE_3rd}
\mu(\mathcal{E})=\mu(L_1^{-1}L_2)>0\quad \Rightarrow \quad d_{ijk}(-m^i+n^i)a^ja^k>0\ .
\end{align}
Putting this all together, three necessary conditions for the slope-stability of ${V_{\rep 3}^\prime}^{(3)}$ are
\begin{equation}
\mu(\mathcal{F})<0\ ,\quad \mu(\mathcal{K})>0\ , \quad \mu(\mathcal{E})>0\ .
\end{equation}

Of course, in addition to these three conditions, it is necessary to require that $V_{\rep 3}^\prime$ satisfy the Bogomolov inequality. That is, 
\begin{equation}
\label{eq:bogolomov_3rd}
\int c_2({V_{\rep 3}^\prime}^{(3)})\wedge \omega\geq 0 \quad \Rightarrow \quad d_{ijk}m^i m^j a^k+3d_{ijk}n^i n^j a^k\leq 0\ .
\end{equation}
Note that the form of the Bogomolov inequality is the same as in \eqref{eq:bogomolov_2lines}. Of course, all of the universal physical constraints remain unchanged.

\begin{figure}[t]
   \centering
     \begin{subfigure}[b]{0.46\textwidth}
\includegraphics[width=1.0\textwidth]{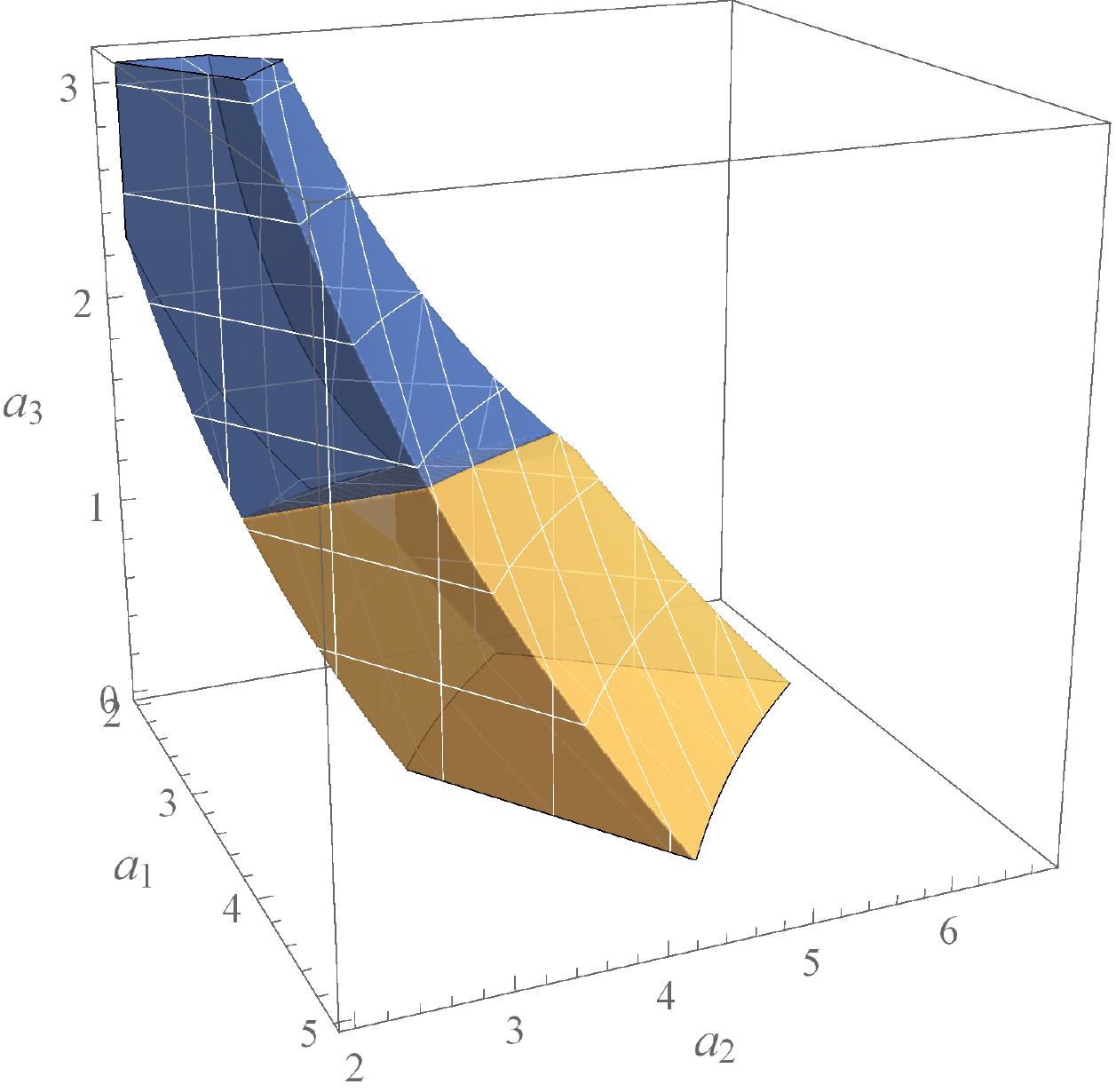}
\end{subfigure}
\caption{ The colored region is the subspace of K\"ahler moduli space which, in addition to satisfying conditions \eqref{51}-\eqref{jack1} outlined in Section \ref{Physical Constraints}, also restricts the strong coupling parameter to the smaller values $\epsilon_S^{\text{eff}} \lesssim 0.2$. Hence, any point in this region also satisfies the linearization constraint \eqref{eq:linearizedCons}. This subspace does not depend on the
hidden sector bundle. The ``orange'' regime is a subspace of this colored region, in which the non-trivial $SU(3)$ bundle $V_{\rep 3}^\prime$, defined as an extension of the two line bundles $L_1=\mathcal{O}_X(-1, -2 ,1)$ and $L_2=\mathcal{O}_X(6, 0, -1)$, can be slope-stable. Note that one can show that the locus where the {\it decomposable} Whitney sum bundle $V_{\rep 3}$ is poly-stable lies outside the colored region shown in Figure \ref{fig:ModuliSolution_orange} and, hence, does not satisfy all of the universal constraints.
}
\label{fig:moduliSolution2}
\end{figure}

Once again, we perform a systematic scan over all pairs of line bundles with $|m^{i}|,|n^{i}|\leq 15$ for $i=1,2,3$. The values of $(a^1,a^2,a^3)$ that we sample sit inside the ``blue'' region of the K\"ahler cone given in Figure \ref{fig:ModuliSolution_orange}. The survey produces multiple solutions, among which are
\begin{equation}
L_1=\mathcal{O}_X(-1,-2, 1)\ , \qquad  L_2=\mathcal{O}_X(6,0, -1)\ .
\label{bird3}
\end{equation}
For this pair of line bundles, we find that the class of the five-brane is
\begin{equation}
W_i=(0,  2, 0)_i\ ,
\end{equation}
which is effective since $W_i\geq 0$ for $i=1,2,3$. Furthermore, the line bundles $L_1^{-1}L_2^{-3}=\mathcal{O}_X(-19,-2,4)$ and $L_1L_2^{-3}=\mathcal{O}_X(-17,2,2)$, associated with the cohomologies of the $\rep 1_{1,3}$ and $\rep 1_{-1,3}$ representations respectively, are clearly non-ample. This ensures that the relevant sequence extension cohomologies $H^1(X,L_1^{-1}L_2^{-3})$ and $H^1(X,L_1L_2^{-3})$ are potentially non-vanishing. As discussed above, these cohomologies need to be computed individually to ensure that they are, in fact, non-zero. However, as mentioned earlier, such computations are beyond the scope of the present paper.
 Assuming that the existence of the extension can be proven, we can check the necessary conditions for stability. We sample values $(a^1,a^2,a^3)$ inside the ``blue'' region in Figure \ref{fig:ModuliSolution_orange} and check that the stability criteria derived for this third extension branch, that is, equations \eqref{eq:slope_3rd}, \eqref{eq:slopeK_3rd} and \eqref{eq:bogolomov_3rd}, are satisfied. The subspace of the ``blue'' region where this extended bundle is potentially slope-stable is displayed in ``orange'' in Figure \ref{fig:moduliSolution2}.

There are other pairs of bundles, in addition to \eqref{bird3}, that satisfy all constraints when we deform the decomposable bundle along the third branch. We display the full results of our survey in Table \ref{tab:full_scan}, along with the analogous solutions for the remaining four branches as well. Again, whether or not each of these solutions exists depends on the pair of $H^{1}$ cohomologies associated with the extension sequences being non-vanishing.

\subsection{Full Branch Structure Scan}\label{sec:full_scan}

In Sections \ref{sec:first_survey} and \ref{sec:different_branches} we derived the sets of constraints for the first and the third extension branches, respectively. The constraints for the remaining four branches can be derived in similar fashion and so we will not give any details. Instead, in Table \ref{tab:branches_stability} we give the necessary conditions on the slopes of the bundles $\mathcal F$, $\mathcal K$ and $\mathcal E$ for each branch. The columns in this table indicate the required signs of the slopes. For example, when we move away from the decomposable locus along the first branch, both $\mathcal{F}$ and $\mathcal{K}$ must have negative slope for the resulting $SU(3)$ bundle to be stable. 
 
 \begin{table}[t!]
	\noindent \begin{centering}
		\begin{tabular}{cccc}
			\toprule 
			Branch &$\mu(\mathcal{F})$ & $\mu(\mathcal{K})$ & $\mu(\mathcal{E}) $ \tabularnewline
									\midrule
			\midrule 
			1& $-$&  $-$  &  $+$\tabularnewline
			\midrule 
			 2& $+$&  $+$  &  $-$\tabularnewline
			 \midrule 
			 3& $-$&  $+$  &  $+$\tabularnewline
			 \midrule 
			 4& $+$&  $-$  &  $-$\tabularnewline
			 \midrule 
			 5& $-$&  $+$  &  $-$\tabularnewline
			 \midrule 
			 6& $+$&  $-$  &  $+$\tabularnewline
			\bottomrule
		\end{tabular}
		\par\end{centering}
		\caption{Necessary stability conditions for the slopes of $\mathcal{F}$, $\mathcal{K}$ and the $\mathcal{E}$.}
		\label{tab:branches_stability}
\end{table}

We have already given the set of solutions for the first branch in Table \ref{tab:solutions_branch1}. We now perform similar scans for each of the remaining extension branches and display the results of the scan in Table \ref{tab:full_scan}. The first two columns of Table \ref{tab:full_scan} show the pairs of line bundles $L_1$ and $L_2$ which satisfy the necessary stability conditions inside the ``blue'' region shown in Figure \ref{fig:ModuliSolution_orange}. We show only those bundles with $|m^i|,|n^i|\leq10$ to save space\footnote{Note that in Sections 3 and 4 we scanned over the larger intervals $|m^i|,|n^i|\leq15$. Hence, for example, the last two entries in Table \ref{tab:solutions_branch1} do not appear in Table \ref{tab:full_scan}.}.
In the third and fourth columns of Table \ref{tab:full_scan} we give the line bundles associated with the fields $C$ and $\tilde C$ which get VEVs in the effective theory. Therefore, the cohomology groups $H^1$ which are associated with these bundles should be non-zero to allow for a non-trivial extension bundle. 
 \begin{table}[t!]
	\noindent \begin{centering}
		\begin{tabular}{c|cc|cc|}
		\Xhline{2\arrayrulewidth}
	 Branch~1 &$L_1$ & $L_2$ &  $\mathcal{F}\otimes \mathcal{E}^*$  &$\mathcal{K}\otimes \mathcal{E}^*$   \\
	 \cline{2-5} 
	&$(-5,-4,2)$ & $(-2,-2,0)$&$(-11,2,2)$  &$(-10,-8,4)$ \\
	\cline{2-5} 
	&$(-5,-1,1)$ & $(2,1,-1)$&$(-11,-4,4)$ & $(-10,-2,2)$\\
	\cline{2-5} 
	\noalign{\vskip 1mm}
	\Xhline{2\arrayrulewidth}
	 Branch~2 &$L_1$ & $L_2$ &  $\mathcal{E}\otimes \mathcal{F}^*$ &  $\mathcal{E}\otimes \mathcal{K}^*$   \\
	\cline{2-5} 
&	$(1,5,-1)$ & $(-2,-1,1)$& $(-7,-8,4)$  & $(-2,-10,2)$ \\
	\cline{2-5} 
		&	$(5,1,-1)$ & $(-2,-1,1)$&$(-11,-4,4)$  & $(-10,-2,2)$\\
	\cline{2-5} 
		&	$(5,4,-2)$ & $(-2,2,0)$&$(-11,2,2)$  & $(-10,-8,4)$\\
	\cline{2-5} 
		\noalign{\vskip 1mm}
	\Xhline{2\arrayrulewidth}
		Branch~3 & $L_1$ & $L_2$ &  $\mathcal{F}\otimes \mathcal{E}^*$    & $\mathcal{F}\otimes \mathcal{K}^*$     \\
	\cline{2-5} 
	&$(-2,-1,1)$ & $(6,0,-1)$&$(-20,-1,4)$  &  $(-16,1,2)$ \\
	\cline{2-5}
		&$(-1,-8,4)$ & $(0,6,0)$&$(-1,-26,4)$   & $(1,-10,-4)$ \\
	\cline{2-5}
			&$(-1,-2,1)$ & $(6,0,-1)$&$(-19,-2,4)$  & $(-17,2,2)$ \\
	\cline{2-5}
		&$(1,2,-1)$ & $(6,0,-1)$&$(-17,2,2)$ & $(-19,-2,4)$\\
	\cline{2-5}
			&$(1,8,-4)$ & $(0,6,0)$&$(1,-10,-4)$ & $(-1,-26,4)$\\
	\cline{2-5}
		&$(2,1,-1)$ & $(6,0,-1)$&$(-16,1,2)$ & $(-20,-1,4)$\\
	\cline{2-5}
		\noalign{\vskip 1mm}
	\Xhline{2\arrayrulewidth}
			Branch~4 & $L_1$ & $L_2$ &  $\mathcal{E}\otimes \mathcal{F}^*$  &  $\mathcal{K}\otimes \mathcal{F}^*$    \\
	\cline{2-5} 
	&$(-2,-1,1)$ & $(-6,0,1)$&$(-16,1,2)$  & $(-20,-1,4)$\\
	\cline{2-5}
		&$(-1,-8,4)$ & $(0,-6,0)$&$(1,-10,-4)$  & $(-1,-26,4)$\\
	\cline{2-5}
	&$(-1,-2,1)$ & $(-6,0,1)$&$(-17,2,2)$ & $(-19,-2,4)$\\
	\cline{2-5}
	&$(1,2,-1)$ & $(-6,0,1)$&$(-19,-2,4)$ & $(-17,2,2)$\\
	\cline{2-5}
	&$(1,8,-4)$ & $(0,-6,0)$&$(-1,-26,4)$  & $(1,-10,-4)$\\
	\cline{2-5}
	&$(2,1,-1)$ & $(-6,0,1)$&$(-20,-1,4)$  & $(-16,1,2)$\\
	\cline{2-5}
		\noalign{\vskip 1mm}
	\Xhline{2\arrayrulewidth}
			Branch~5 & $L_1$ & $L_2$ &  $\mathcal{E}\otimes \mathcal{K}^*$ &  $\mathcal{F}\otimes \mathcal{K}^*$   \\
	\cline{2-5} 
	&$(1,-5,-1)$ & $(2,1,-1)$&$(-2,-10,2)$  & $(-7,-8,4)$\\
	\cline{2-5}
	&$(5,1,-1)$ & $(2,1,-1)$&$(-10,-2,2)$ & $(-11,-4,4)$\\
	\cline{2-5}
	&$(5,4,-2)$ & $(2,-2,0)$&$(-10,-8,4)$ &  $(-11,2,2)$\\
	\cline{2-5}
		\noalign{\vskip1mm}
	\Xhline{2\arrayrulewidth}
				Branch~6 & $L_1$ & $L_2$ &  $\mathcal{K}\otimes \mathcal{E}^*$ &   $\mathcal{K}\otimes \mathcal{F}^*$    \\
	\cline{2-5} 
	&$(-5,-4,2)$ & $(-2,2,0)$&$(-10,-8,4)$ &  $(-11,2,2)$\\
	\cline{2-5}
	&$(-5,-1,1)$ & $(-2,-1,1)$&$(-10,-2,2)$ & $(-11,-4,4)$\\
	\cline{2-5}
	&$(-2,-10,1)$ & $(1,2,-3)$&$(-4,-20,2)$ & $(1,-4,-8)$\\
	\cline{2-5}
	&$(-1,-5,1)$ & $(-2,-1,1)$&$(-2,-10,2)$ & $(-7,-8,4)$\\
	\cline{2-5}
		\end{tabular}
		\par\end{centering}
		\caption{In the first two columns, we list pairs of line bundles $L_1$ and $L_2$ which satisfy the necessary stability conditions inside the ``blue'' region shown in Figure \ref{fig:ModuliSolution_orange}. In the third and forth columns, we present the line bundles associated with the charged matter fields which get VEVs. Note that we have used an abbreviated notation to represent the line bundles, $\mathcal{O}_X(a,b,c)\equiv (a,b,c)$.}
		\label{tab:full_scan}
\end{table}
Showing that any of these extensions is non-zero would encourage a full check of stability for the relevant extension bundle, as they would give a plausible hidden sector for the $B-L$ MSSM.
Compared with the construction of hidden sectors from a single line bundle presented in \cite{Ashmore:2020ocb}, we have many more possible solutions when using two line bundles. 

\section{Conclusion}

In this paper, we attempted to build a hidden sector bundle using two line bundles together with an explicit embedding into $E_{8}$, via the two line bundle vectors given in \eqref{eq:u1_vectors} or, equivalently, by embedding them into the $SU(3)$ factor of the maximal subgroup $E_{6} \times SU(3)$. It should be clear, however, that the analysis and methods developed in this paper can be applied to models with hidden sectors built from different embeddings into $E_{8}$. For example,
consider two line bundles embedded into $E_{8}$, not with the generators given in \eqref{eq:u1_vectors}, but, rather, with two different linearly independent generators 
\begin{align}
	\boldsymbol{t}_{1} & =(0,0,0,0,0,0,0,-2)\ ,\\
	\boldsymbol{t}_{2} & =(0,0,0,0,0,0,-2,0)\ ,
\end{align}
with the two different line bundle vectors given by
\begin{equation}
\boldsymbol{V}_{i}=m^{i}\boldsymbol{t}_{1}+n^{i}\boldsymbol{t}_{2}\ \quad i=1,2,3\ .
\end{equation}
The non-Abelian commutant of $U(1) \times U(1)$ inside $E_{8}$ is then $SO(12)$. Using the \emph{Mathematica} package LieART~\cite{Feger:2012bs,Feger:2019tvk}, one can then find the decomposition of the adjoint representation of $\Ex 8$ with respect to $SO(12) \times U(1) \times U(1)$. It is given by
\begin{equation}
	\label{eq:decompSO12big}
	\begin{split}
		\rep{248}&=\rep{66}_{0,0}+2\times\rep{1}_{0,0}+\rep{1}_{2,0}+\rep{1}_{-2,0}+\rep{1}_{0,2}+\rep{1}_{0,-2}\\
		&\eqspace+\rep{32}_{1,0}+\rep{32}_{-1,0}+\rep{32}_{1,1}+\rep{32}_{1,-1}\\
		&\eqspace+\rep{12}_{1,1}+\rep{12}_{-1,1}+\rep{12}_{1,-1}+\rep{12}_{-1,1}\ .
	\end{split}
\end{equation}
More prosaically, this breaking pattern can be obtained by first breaking $E_8$ to $E_7\times SU(2)$, under which the adjoint representation of $E_8$ decomposes as
\begin{equation}
\label{eq:SO12first}
\rep{248}=(\rep{133},\rep{1})+(\rep{56},\rep{2})+(\rep{1},\rep{3})\ .
\end{equation}
Breaking $E_7$ further to $SO(12)\times SU(2)$, the $E_7$ representations that appear above decompose as
\begin{equation}
\label{eq:SO12second}
\begin{split}
\rep{133}&=(\rep{66},\rep{1})+(\rep{32},\rep{2})+(\rep{1},\rep{3})\ ,\\
\rep{56}&=(\rep{12},\rep{2})+(\rep{32},\rep{1})\ .\\
\end{split}
\end{equation}
Finally, we break both $SU(2)$ groups down to $U(1)$s, with the fundamental representation of $SU(2)$ decomposing as $\rep{2}=\rep{1}_{1}+\rep{1}_{-1}$. Putting this together, one sees that the adjoint representation of $E_8$ decomposes under $SO(12)\times U(1)\times U(1)$ exactly as in \eqref{eq:decompSO12big} above.

Each of the two line bundles $L_1$ and $L_2$ associated with the two $U(1)$ factors are embedded into a different $SU(2)$ bundle. At the decomposable locus, we have a reducible rank-four bundle
\begin{equation}
\label{eq:defineBundlesSU222}
\begin{split}
V_{\rep2,\rep2}=( L_1^{-1}\oplus L_1)\oplus( L_2^{-1}\oplus L_2)\ ,
\end{split}
\end{equation}
with $U(1)\times U(1)$ structure group. In this case, the only singlets under the non-Abelian $SO(12)$ factor that are charged under the two $U(1)$s are $\rep 1_{2,0}$, $\rep 1_{0,2}$ and their conjugates. These give charged singlet matter fields whose VEVs may be used to set the D-terms associated with the low-energy $U(1)$s to zero. Turning on a combination of two VEVs and, hence, extending the associated exact sequences, would correspond to deforming the hidden sector bundle with $U(1)\times U(1)$ structure group into one with $\SU 2\times \SU 2$ instead. If each $SU(2)$ bundle can be made slope-stable, the extended version of $V_{\rep2,\rep2}$ will then be slope poly-stable.

We conclude that there are clearly a large number of different ways of consistently embedding two line bundles into $E_{8}$ and, hence, many different hidden sector bundles for the $B-L$ MSSM that can be constructed in this manner. It is also clear that there are a large number of hidden sectors that one can construct using three or more line bundles with various embeddings into $E_{8}$. We hope to come back to this in future work.

\subsubsection*{Acknowledgements}

AA is supported by the European Union's Horizon 2020 research and innovation program under the Marie Sk\l{}odowska-Curie grant agreement No.~838776. SD is supported in part by research grant DOE No.~DESC0007901. BAO is supported in part by the research grant DOE No.~DESC0007901 and SAS Account 020-0188-2-010202-6603-0338.

\appendix

\section{Conventions}\label{sec:Conventions}

\begin{figure}[t]
	\centering
	\includegraphics[width=0.9\textwidth]{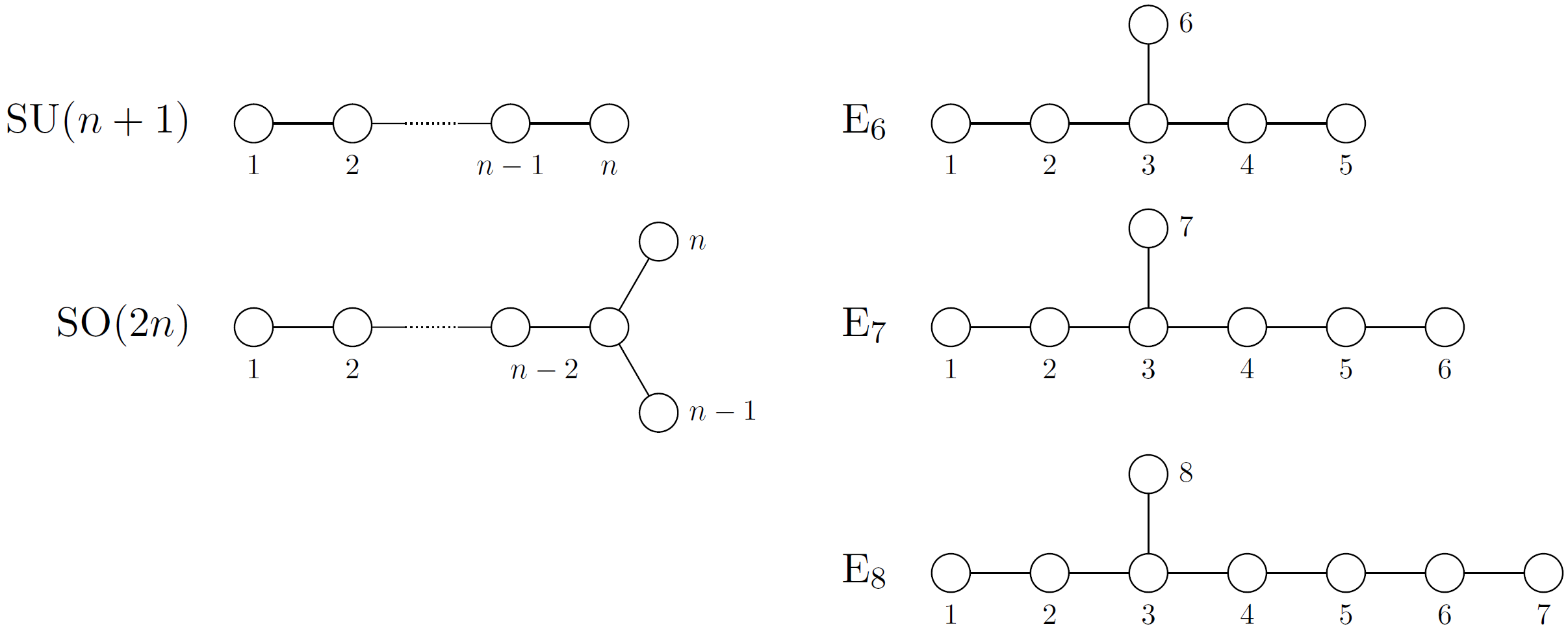}
	\caption{Dynkin diagrams of the classical Lie groups. We follow the Dynkin convention for labelling simple roots, as used in \cite{Feger:2012bs,Feger:2019tvk}.}
	\label{fig:e8}
\end{figure}

We follow the Dynkin convention for labelling the simple roots of $\ex 8$, in agreement with the \emph{Mathematica} package LieART~\cite{Feger:2012bs,Feger:2019tvk} which we used in many of the line bundle vector calculations. In particular, we number the nodes of the Dynkin diagram as in Figure \ref{fig:e8}. We mostly work in the ``orthogonal basis'' $\{e_{a}\}$ with $a=1,\ldots,8$ (see \cite{Feger:2012bs} for more details), where the components of the root vectors are given with respect to an orthogonal basis. In particular, the line bundle vectors $\boldsymbol{V}_{i}$ and roots $\boldsymbol{r}$ in the main text are expressed in the orthogonal basis. The eight simple roots $\alpha_{I}=\alpha_{I}^{a}e_{a}$ of $\ex 8$ are given in this basis by
\begin{equation}
	\begin{aligned}\alpha_{1}^{a} & =\tfrac{1}{2}(1,-1,-1,-1,-1,-1,-1,1)\ , & \alpha_{5}^{a} & =(0,0,0,-1,1,0,0,0)\ ,\\
		\alpha_{2}^{a} & =(-1,1,0,0,0,0,0,0)\ , & \alpha_{6}^{a} & =(0,0,0,0,-1,1,0,0)\ ,\\
		\alpha_{3}^{a} & =(0,-1,1,0,0,0,0,0)\ , & \alpha_{7}^{a} & =(0,0,0,0,0,-1,1,0)\ ,\\
		\alpha_{4}^{a} & =(0,0,-1,1,0,0,0,0)\ , & \alpha_{8}^{a} & =(1,1,0,0,0,0,0,0)\ .
	\end{aligned}
	\label{eq:roots}
\end{equation}

In addition to the orthogonal basis we also have the $\alpha$-basis and the $\omega$-basis. The $\alpha$-basis is the basis of simple roots. This has the advantage that it shows precisely how a given root is made from a sum of simple roots. In this basis, the components of the simple roots are given by
\begin{equation}
	\tilde{\alpha}_{1}^{a}=(1,0,0,0,0,0,0,0)\ ,\qquad\tilde{\alpha}_{2}^{a}=(0,1,0,0,0,0,0,0)\ ,
\end{equation}
and so on, so that $\alpha_{I}=\tilde{\alpha}_{I}^{a}\alpha_{a}=\delta_{I}^{a}\alpha_{a}$. Finally, the $\omega$-basis is the basis of fundamental weights, also known as the Dynkin basis. This basis is such that the simple roots correspond to the rows of the Cartan matrix $A_{ab}$ (for $\ex 8$ in this case), so that, for example, the first two simple roots can be written as
\begin{equation}
	\hat{\alpha}_{1}^{a}=(2,-1,0,0,0,0,0,0)\ ,\qquad\hat{\alpha}_{2}^{a}=(-1,2,-1,0,0,0,0,0)\ ,
\end{equation}
where we have written the simple roots in the $\omega$-basis as $\alpha_{I}=\hat{\alpha}_{I}^{a}\omega_{a}$. Note that for algebras (such as $\ex 8$) whose roots are all of length 2, the $\alpha$-basis and $\omega$-basis are dual
\begin{equation}
	(\alpha_{a},\omega_{b})=\delta_{ab}\ .
\end{equation}
In particular, this implies
\begin{equation}
	(\alpha_{I},\alpha_{J})=\tilde{\alpha}_{I}^{a}\hat{\alpha}_{J}^{b}(\alpha_{a},\omega_{b})=\delta_{I}^{a}\hat{\alpha}_{J}^{b}\delta_{ab}=A_{IJ}\ .
\end{equation}
The transformations between these bases are given by
\begin{equation}
	\alpha_{a}=\sum_{b}A_{ab}\omega_{b}\ ,\qquad\omega_{a}=\sum_{b}\hat{\Omega}_{ab}e_{b}\ ,
\end{equation}
where $\hat{\Omega}_{ab}$ is the matrix whose rows are the fundamental weights in the orthogonal basis and $A_{ab}$ is the Cartan matrix of $\ex 8$, given by
\begin{equation}
	A_{ab}=\left(\begin{array}{cccccccc}
		2 & -1 & 0 & 0 & 0 & 0 & 0 & 0\\
		-1 & 2 & -1 & 0 & 0 & 0 & 0 & 0\\
		0 & -1 & 2 & -1 & 0 & 0 & 0 & -1\\
		0 & 0 & -1 & 2 & -1 & 0 & 0 & 0\\
		0 & 0 & 0 & -1 & 2 & -1 & 0 & 0\\
		0 & 0 & 0 & 0 & -1 & 2 & -1 & 0\\
		0 & 0 & 0 & 0 & 0 & -1 & 2 & 0\\
		0 & 0 & -1 & 0 & 0 & 0 & 0 & 2
	\end{array}\right)\ .
\end{equation}

Given the transformations between the various bases, it is simple to write down line bundle vectors that break particular combinations of the simple roots. Consider the inner product of the line bundle vector $V$ with the $I^{\text{th}}$ simple root $\alpha_{I}$:
\begin{equation}
	V\cdot\alpha_{I}=\hat{V}^{a}\tilde{\alpha}_{I}^{b}(\omega_{a},\alpha_{b})=\hat{V}^{a}\delta_{I}^{b}\delta_{ab}=\hat{V}_{I}\ .
\end{equation}
This means that the inner product of a line bundle vector $V$ with a simple root $\alpha_{I}$ is given by the $I^{\text{th}}$ component of the line bundle vector written in the $\omega$-basis. If, for example, we want to pick a line bundle vector that breaks the first simple root of $\ex 8$ and is orthogonal to the others, we can take $\hat{V}^{a}=(1,0,0,0,0,0,0,0)$. Transforming back to the orthogonal basis (which is the basis used in the main text), we would then have
\begin{equation}
	\boldsymbol{V}=(0,0,0,0,0,0,0,2)\ .
\end{equation}
This would lead to an unbroken $SO(14)$ group.

\section{Anomaly Cancellation}\label{app:anomaly_cancellation}

As discussed in~\cite{Lukas:1998tt,Donagi:1998xe}, anomaly cancellation in heterotic M-theory requires that
\begin{equation}
  c_2(TX)-c_2(\mathcal{V}^{(1)})
  -\op{ch}_2(\mathcal{V}^{(2)}) - W 
  = 0 \ ,
  \label{29}
\end{equation}
where $\mathcal{V}^{(1)}$ is the observable $SU(4)$ bundle, $\mathcal{V}^{(2)}$ is the hidden sector bundle -- whose composition is the main interest in this paper -- $TX$ is the tangent bundle of the compactification threefold, while $W$ is the effective class of the single five-brane between the hidden and observable sector.
Using results for the second Chern class of the observable sector tangent bundle and gauge bundle of the $B-L$ MSSM given in \cite{Ovrut:2018qog}, we find that
\begin{equation}
  \frac{1}{v^{1/3}}\int_X \left(c_2(TX)-c_2(\mathcal{V}^{(1)})\right)\wedge \omega_i=\left( \tfrac{4}{3},\tfrac{7}{3},-4\right)_i\ .
\end{equation}
Similarly, it was shown in \cite{Ovrut:2018qog} that for a generic hidden sector vector bundle of the form
\begin{equation} 
\mathcal{V}^{(2)}= \mathcal{V}_{N} \oplus \mathcal{L} , \quad  \mathcal{L}= \bigoplus_{r=1}^{R} L_{r}
\label{sun1}
\end{equation}
where $\mathcal{V}_{N}$ is a slope-stable, non-Abelian bundle and each $L_{r}=\mathcal{O}_{X}(l_{r}^1,l_{r}^2,l_{r}^3)$ is a holomorphic line bundle with structure group U(1), that
\begin{equation}
  \frac{1}{v^{1/3}}\int_X \left(-\rm{ch}_2(\mathcal{V}^{(2)})\right)\wedge \omega_i= -d_{ijk}c_{N}^{jk}+ \sum_{r=1}^{R} a_{r}d_{ijk} l_{r}^{j}l_{r}^{k} \ .
\end{equation}
The coefficients $d_{ijk}$ are the intersection numbers associated with our specific Schoen threefold. They were given in \cite{Ashmore:2020wwv,Ovrut:2018qog}, for example. However, since they are used extensively in this paper, we present them here again for completeness. They are given by
\begin{equation}\label{4new}
  (d_{ijk}) = 
  \left(
    \begin{array}{ccc}
      \(0,\tfrac13,0\) & \(\tfrac13,\tfrac13,1\) & \(0,1,0\) \\
      \(\tfrac13,\tfrac13,1\) & \(\tfrac13,0,0\) & \(1,0,0\) \\
      \(0,1,0\) & \(1,0,0\) & \(0,0,0\)
    \end{array} 
  \right) .
\end{equation}
The $(i,j)$-th entry in the matrix corresponds to the triplet
$(d_{ijk})_{k=1,2,3}$.
The coefficient 
\begin{equation}
a_{r}=\frac{1}{4} \tr Q_{r}^{2} \ ,
\label{sun3}
\end{equation}
with $Q_{r}$ the generator of the $r$-th $U(1)$ factor embedding into the $\underline{\bf{248}}$ representation of the hidden sector $E_{8}$.
If we define 
\begin{equation}
W_{i}= \frac{1}{v^{1/3}}\int_X{W \wedge \omega_{i}} \ ,
\label{sun2}
\end{equation}
it follows that the anomaly condition \eqref{29} can be re-expressed as
\begin{equation}
\label{eq:333}
W_i=\big( \tfrac{4}{3},\tfrac{7}{3},-4\big)\big|_i -d_{ijk}c_{N}^{jk}+ \sum_{r=1}^{R} a_{r}d_{ijk} l_{r}^{j}l_{r}^{k} \ .
\end{equation}
As discussed in \cite{Ovrut:2018qog}, in order for each line bundle $L_{r}$ to arise from a $\mathbb{Z}_{3} \times \mathbb{Z}_{3}$ equivariant line bundle on the covering space of X, the integers $l_{r}^{i}$, $i=1,2,3$ of  $L_{r}=\mathcal{O}_{X}(l_{r}^1,l_{r}^2,l_{r}^3)$ must satisfy that constraint that
\begin{equation}
( l_{r}^{1} + l_{r}^{2})~{\rm{mod}}~3=0 \ .
\label{dark1}
\end{equation}
Furthermore, in order to preserve $N=1$ supersymmetry $W$ must be an effective class; that is, each component $W_{i}$,  $i=1,2,3$ must be non-negative.

In Section \ref{2 line bundles embedding}, we study a hidden sector composed of two line bundles $L_1=\mathcal{O}_X(m^1,m^2,m^3)$ and  $L_2=\mathcal{O}_X(n^1,n^2,n^3)$ embedded into an $SU(3)\subset E_8$ connection with no non-Abelian bundle factor. In this case, the second Chern character of the hidden sector bundle is found to be to be 
\begin{equation}
\begin{split}
\op{ch}_{2}(\mathcal{V}^{(2)})=&\,c_{1}(L_1)\wedge c_{1}(L_1)+3c_{1}(L_2)\wedge c_{1}(L_2)\\
&=\frac{1}{v^{2/3}}(m^1\omega_{1}+m^2\omega_{2}+m^3\omega_{3})^{2}+\frac{3}{v^{2/3}}(n^1\omega_{1}+n^2\omega_{2}+n^3\omega_{3})^{2}\ .
\end{split}
\end{equation}
The anomaly condition then takes the form
\begin{equation}
\label{eq:anomaly_modified2}
W_i=\big( \tfrac{4}{3},\tfrac{7}{3},-4\big)\big|_i+d_{ijk}m^jm^k+3d_{ijk}n^jn^k\ .
\end{equation}

Finally, we note that in our model with a single five brane the $\beta^{(n)}_i $ charges have the form
\begin{align}
  \beta^{(0)}_i &= 
  \big( \tfrac{2}{3},-\tfrac{1}{3},4 \big)\big|_i \ ,\\
  \beta^{(1)}&=W_i\ , \\
\beta^{N+1}_i&=-\beta_i^{(0)}-W_i\ .
  \label{35}
\end{align}
\section{Linearization Constraints}\label{app:linear}

The five-dimensional effective theory of heterotic M-theory, obtained
by reducing Hořava--Witten theory on the Calabi--Yau
threefold, admits a BPS double domain wall solution with five-branes
in the bulk space. This double domain wall was analyzed in in detail in a series of papers \cite{Donagi:1999gc,Lukas:1998tt,Lukas:1998hk,Lukas:1998yy,Lukas:1997fg,Brandle:2001ts}. We will summarize some of the relevant results here, and we will make use of the notation outlined in \cite{Ashmore:2020ocb}. 

The detailed structure of the linearized double domain wall depends on the solution of three non-linear equations discussed in \cite{Lukas:1998tt,Brandle:2003uya}. These can be approximately solved by expanding to linear order\footnote{This set of
non-linear equations was also solved to second order in \cite{Ashmore:2020ocb}.} in the expansion parameter 
\begin{equation}
\epsilon^{\text{eff}}_S=\frac{\epsilon_S^\prime\hat R}{V} \ . 
\label{flow1}
\end{equation}
It was shown in \cite{Lukas:1998tt} that the 
conditions for the validity of the linear approximation then break
into two parts. Written in terms of the averaged moduli, these are
\begin{equation}
  2\epsilon_S'\frac{\Rhat}{V^{1/3}}
  \left|
    \beta_i^{(0)} \big(z-\tfrac{1}{2}\big)
    -\frac{1}{2}W_i(\tfrac{1}{2}-\lambda)^2
  \right|
  \ll 
  \left| d_{ijk} a^j a^k \right|
  , \quad z \in [0,z_1]
\label{45B}
\end{equation}
and 
\begin{equation}
  2\epsilon_S'\frac{\Rhat}{V^{1/3}}
  \left|
    (\beta_i^{(0)}+W_i)
    \big(z-\tfrac{1}{2}\big)
    -\frac{1}{2}W_i(\tfrac{1}{2}+\lambda)^2
  \right| 
  \ll 
  \left| d_{ijk} a^j a^k \right|
  , \quad z \in [z_1,1] .
\label{45C}
\end{equation}
Here we define  $z=\frac{x^{11}}{\pi\rho}$, where $x^{11}$ is the coordinate along the dimension which separates
the observable and the hidden sector in the BPS state, while $\pi \rho$ is the separation parameter between these sectors. In our setup, $z\in [0,1]$. The observable sector and the hidden sector are located at $z=0$ and $z=1$, respectively,
while the five brane is set at $z=\lambda+\tfrac{1}{2}$. In the absence of a mechanism to fix the five-brane position, $\lambda$ is another modulus in our system. For consistency with our previous work\cite{Ashmore:2020ocb}, we fix the position of the five-brane close to the hidden sector, at $\lambda=0.49$.

\section{Genus-One Corrected FI Terms}\label{app:FI}

As is commonly known, a $\Uni 1$ symmetry that appears in the both the internal and four-dimensional gauge generate a D-term potential proportional to an FI term associated with the $U(1)$ bundle~\cite{Dine:1986zy,Dine:1987xk,Lukas:1999nh,Blumenhagen:2005ga}
The expression for the genus one corrected FI term associated with a $U(1)$ bundle $L$ on the hidden sector was computed in \cite{Weigand:2006yj,Blumenhagen:2005ga} within the context of the weakly coupled heterotic string. It was then shown in \cite{Ovrut:2015uea} that in the strongly coupled limit this expression becomes
\begin{equation}
\label{eq:B1_App}
FI_{L}=\frac{a_L}{2}\frac{\epsilon_S\epsilon_R^2}{\kappa_4^2}\frac{1}{\hat R V^{2/3}}\left[\mu(L)+\frac{\epsilon_S^\prime \hat R}{V^{1/3}}\int_X c_1(L)\wedge \left(  J^{(N+1)}+\sum^N_{n=1}z_n^2J^{(n)}  \right)\right]\ ,
\end{equation} 
where the complex two-forms $J$ are defined in  \cite{Ovrut:2015uea} and $n$ runs over all five-branes in the bulk interval. In our set up, we only have one five brane at position $z=\lambda +\frac{1}{2}$  with the source term given by $J^1=W$. The coefficient $a_L$ depends on the exact embedding of the line bundle $L$ associated with the FI term into the hidden sector $E_8$. 
In the case of a hidden sector with a single line bundle, there is one FI term associated with it. In this case, the coefficient $a_L$ is simply equal to the coefficient $a$ derived for the second Chern character. For the embedding $U(1)\rightarrow SU(2)\rightarrow E_8$,
we found in \cite{Ashmore:2020ocb} that $a_L=1$.

In the case of a hidden sector with two line bundles, we have two $FI$ terms. Each is  associated with one of the two line bundles $\mathcal{F}$ and $\mathcal{K}$ defined in the decomposition 
\begin{equation}
V_{\rep 3}=\mathcal{F}\oplus \mathcal{K} \oplus \mathcal{E}
\end{equation}
of the $SU(3)$ bundle $V_{\rep 3}$ at the decomposable locus.
Note that the line bundle $\mathcal{E}$ depends on $\mathcal{K}$ and $\mathcal{F}$ such that 
$c_1(\mathcal{E})=-c_1(\mathcal{F})-c_1(\mathcal{K})$. Hence, $V_{\rep 3}$ has the structure group
$S(U(1)\times U(1)\times U(1))\sim U(1)\times U(1)$ at the stability wall. The genus-one corrected FI terms in this case are
\begin{equation}
\begin{split}
\label{eq:B3_App}
&FI_{\mathcal{F}}=\frac{a_{\mathcal{F}}}{2}\frac{\epsilon_S\epsilon_R^2}{\kappa_4^2}\frac{1}{\hat R V^{2/3}}\left[\mu(\mathcal{F})+\frac{\epsilon_S^\prime \hat R}{V^{1/3}}\int_X c_1(\mathcal{F})\wedge \left(  J^{(N+1)}+\sum^N_{n=1}z_n^2J^{(n)}  \right)\right]\ , \\
&FI_{\mathcal{K}}=\frac{a_{\mathcal{K}}}{2}\frac{\epsilon_S\epsilon_R^2}{\kappa_4^2}\frac{1}{\hat R V^{2/3}}\left[\mu(\mathcal{K})+\frac{\epsilon_S^\prime \hat R}{V^{1/3}}\int_X c_1(\mathcal{K})\wedge \left(  J^{(N+1)}+\sum^N_{n=1}z_n^2J^{(n)}  \right)\right] \ .
\end{split}
\end{equation} 
Note that the $FI$ terms in the effective theory are associated with the line bundles $\mathcal{F}=L_2^{-2}$ and $\mathcal{K}=L_1L_2$ and \emph{not} with the bundles $L_1=\mathcal{O}_X(m^1,m^2,m^3)$ and $L_2=\mathcal{O}_X(n^1,n^2,n^3)$. 
Hence, the coefficients $a_{\mathcal{F}}$ and $a_{\mathcal{K}}$ in front of the expressions in \eqref{eq:B3_App} depend on the how the bundles 
$\mathcal{F}$ and $\mathcal{K}$ embed into the $E_8$ connection. This calculation was trivial in the single bundles case, because we parametrized directly all the equations in terms of the bundle $L$ associated with the $FI$ term. For the two line bundle case of interest in this paper, one can read off the generators $Q_{\mathcal{F}}$ and $Q_{\mathcal{K}}$ associated with their embedding into $E_8$ from Table \ref{tab:4thTable}. These are given by
\begin{align}
&Q_{\mathcal{F}}=(1,-1,2,-2,-1,1,-\id_{27},0\times \id_{27}, \id_{27},\id_{27},0\times \id_{27}, -\id_{27})\ ,\\
&Q_{\mathcal{K}}=(-1,1,1,-1,-2,2,0\times \id_{27},- \id_{27}, \id_{27},0\times\id_{27},\id_{27}, -\id_{27})\ .
\end{align}
Hence, we get
\begin{align}
&a_{\mathcal{F}}=\frac{1}{4} \tr Q_{\mathcal{F}}^2=1,\\
&a_{\mathcal{K}}=\frac{1}{4} \tr Q_{\mathcal{K}}^2=1\ .
\end{align}

As a side note, one might realize that we could have parametrized our equations in terms of the bundles $\mathcal{F}$ and $\mathcal{K}$ directly and write 
\begin{equation}
\mathcal{F}=\mathcal{O}_X(p^1,p^2,p^3)\ , \quad \mathcal{K}=\mathcal{O}_X(r^1,r^2,r^3)\ ,
\end{equation}
such that
\begin{equation}
\mathcal{E}=\mathcal{O}_X(-(p^1+r^1),-(p^2+r^2),-(p^3+r^3))\ .
\end{equation}
This description is related to the one we have used by the transformations
\begin{equation}
p^i=-2m^i\ ,\quad  r^i=m^i+n^i\ ,\quad i=1,2,3\ .
\end{equation}
When using this description, however, the constraint equations derived in the main text become considerably more complicated. In particular,
note that $\tr Q_{\mathcal{F}}Q_{\mathcal{K}}\neq 0$. This non-zero mixing term leads to more convoluted expressions for the second Chern character $\op{ch}(\mathcal{V}^{(2)})$ and any equations that contain it, such as the anomaly condition. Of course, the two descriptions are equivalent and lead to the same results in the end. However, we will {\it not} use this second parametrization in this paper.

For the two line bundles system with $L_1=\mathcal{O}_X(m^1,m^2,m^3)$ and $L_2=\mathcal{O}_X(n^1,n^2,n^3)$, we obtain
\begin{equation}
\begin{split}
FI_{\mathcal{F}}=\frac{1}{2}\frac{\epsilon_S\epsilon_R^2}{\kappa_4^2}\frac{1}{\hat R V^{2/3}}&\left[-2d_{ijk}n^ia^ja^k+\frac{2\epsilon_S^\prime \hat R}{V^{1/3}}n^i \Big(   \big( \tfrac{2}{3},-\tfrac{1}{3},4 \big)\big|_i  +\left(1-(1+\tfrac{\lambda}{2})^2\right) W_i    \Big)\right]\ ,\\
FI_{\mathcal{K}}=\frac{1}{2}\frac{\epsilon_S\epsilon_R^2}{\kappa_4^2}\frac{1}{\hat R V^{2/3}}&\bigg[d_{ijk}(m^j+n_j)^ia^ja^k
\\&-\frac{\epsilon_S^\prime \hat R}{V^{1/3}}(m^i+n^i) \Big(   \big( \tfrac{2}{3},-\tfrac{1}{3},4 \big)\big|_i  +\left(1-(1+\tfrac{\lambda}{2})^2\right) W_i    \Big)\bigg]\ .\\
\end{split}
\end{equation}
$W_i$ is given in eq. \eqref{eq:anomaly_modified2} for a system with two line bundles embedded into $SU(3)$. 

\section{Gauge Threshold Corrections}\label{app:gauge_couplings}

The gauge couplings of the non-anomalous components of the $d=4$ gauge
group, in both the observable and hidden sectors, have been computed
to order $\kappa_{11}^{4/3}$ in~\cite{Lukas:1998hk}. Written in terms of the K\"ahler moduli $a^{i}$, these are given by
\begin{equation}
  \frac{4\pi}{(g^{(1)})^2} \propto
  V (1+\epsilon_S' \frac{\Rhat}{2V^{4/3}} 
   \sum_{n=0}^{N}(1-z_n)^2 a^i \beta^{(n)}_i )
\label{62}
\end{equation}
and 
\begin{equation}
  \frac{4\pi}{(g^{(2)})^2} \propto
  V (1+\epsilon_S' \frac{\Rhat}{2V^{4/3}} 
  \sum_{n=1}^{N+1}z_n^2 a^i\beta^{(n)}_i)
\label{63}
\end{equation}
respectively. The positive definite constant of proportionality is identical for both gauge couplings and is not relevant to the present discussion. It is important to note that the effective parameter of the $\kappa_{11}^{2/3}$ expansion is, as discussed above,  $\epsilon_{S}^{\rm eff}=\epsilon_S' \frac{\Rhat}{V}$.

Consistency of the $d=4$ effective theory requires both
$(g^{(1)})^2$ and $(g^{(2)})^2$ to be positive. Expressing $V=\tfrac{1}{6}d_{ijk}a^ia^ja^k$ and the $\beta^n$ charges as in eq \eqref{35}, we can write these two constraints in terms of the K\"ahler moduli:
\begin{align}
 (g^{(1)})^2 > 0 \>\Rightarrow\> &d_{ijk}a^ia^ja^k+3\frac{\epsilon_S^\prime \hat R}{V^{1/3}}\big(\tfrac{2}{3}a^1-\tfrac{1}{3}a^2+4a^3+(\tfrac{1}{2}-\lambda)^2W_ia^i\big)>0 \ ,\\
  (g^{(2)})^2 >0 \>\Rightarrow\>   &d_{ijk}a^ia^ja^k-3\frac{\epsilon_S^\prime \hat R}{V^{1/3}}\big(\tfrac{2}{3}a^1-\tfrac{1}{3}a^2+4a^3+(1-(\tfrac{1}{2}+\lambda))^2W_ia^i\big)>0 \ .
\end{align}
Changing the model we use for the hidden sector bundle is reflected solely in the expression for $W_i$. $W_i$ is given in eq \eqref{eq:333} for a generic hidden sector vector bundle of the form \eqref{sun1}.

\section{Subbundles of Isomorphic Extension Bundles}\label{app:subbundles}

In Table 3, we presented the six different extension  branches for deforming the Whitney sum $V_{\rep 3}=\mathcal{F}\oplus\mathcal{K}\oplus\mathcal{E}$ away 
from the decomposable locus. For each such branch, there are two different pairs of sequences which lead, however, to isomorphic $SU(3)$ bundles. Because of this isomorphism, in Table 3 we presented only one pair of these sequences.  Here, however, we need to discuss both of them. For simplicity, let us restrict the discussion to the {\it first} extension branch only. However, the conclusions will apply to the remaining five branches as well. Let us briefly review the two sets of extension sequences in the first extension branch. These are
\begin{equation}
\begin{split}
\label{eq:sequences_app1}
\op{Ext}^1(\mathcal{E},\mathcal{F})=H^1(X,\mathcal{F}\otimes \mathcal{E}^*)\neq 0\quad  \Rightarrow\quad  &0\rightarrow \mathcal{F}\rightarrow W\rightarrow \mathcal{E}\rightarrow  0\ ,\\\op{Ext}^1(W,\mathcal{K})=H^1(X,\mathcal{K}\otimes \mathcal{E}^*) \neq 0\quad \Rightarrow \quad &0\rightarrow \mathcal{K}\rightarrow V^{\prime}_{\rep 3} \rightarrow W \rightarrow 0\ ,
\end{split}
\end{equation}
or
\begin{equation}
\begin{split}
\label{eq:sequences_app2}
\op{Ext}^1(\mathcal{E},\mathcal{K})=H^1(X,\mathcal{K}\otimes \mathcal{E}^*)\neq 0\quad  \Rightarrow\quad  &0\rightarrow \mathcal{K}\rightarrow W^\prime\rightarrow \mathcal{E}\rightarrow  0\ ,\\
\op{Ext}^1(W^\prime,\mathcal{F})=H^1(X,\mathcal{F}\otimes \mathcal{E}^*) \neq 0\quad \Rightarrow \quad &0\rightarrow \mathcal{F}\rightarrow \mathcal{V}^{\prime}_{\rep 3} \rightarrow W^\prime \rightarrow 0\ .
\end{split}
\end{equation}
However, following the calculation in \cite{Anderson:2010ty}, it can be shown that the resulting $SU(3)$ bundles are actually isomorphic $V^\prime_{\rep 3}\simeq \mathcal{V}_{\rep 3}^\prime$ and so it does not matter which extension one uses.

From the first set definition of the extension in \eqref{eq:sequences_app1}, we learn there is an embedding
\begin{equation}
\mathcal{K}\hookrightarrow V^\prime_{\rep 3}\ ,
\end{equation}
and, hence, $\mathcal{K}$ is a rank-one subbundle of $V_{\rep 3}^\prime$. This means $V_{\rep 3}^\prime$ is stable only if the slope of $\mathcal{K}$ is less than the slope of $V_{\rep 3}^\prime$ (which vanishes):
\begin{equation}
\mu(\mathcal{K})<0\ .
\end{equation}
From the second definition of the extension in \ref{eq:sequences_app2}, we learn that $\mathcal{F}$ is a subbundle of $\mathcal{V}_3^\prime$, which itself is isomorphic to $V_{\rep 3}^\prime$. An obvious question is whether $\mathcal{F}$ is thus a subbundle of $V_{\rep 3}^\prime$ as well, which would then constrain the slope of $\mathcal{F}$ to be negative.

Recall that a sheaf $\mathcal{F}$ is a sub-sheaf of $V$ if it has smaller rank and 
and there exists an embedding $\mathcal{F} \hookrightarrow V$~\cite{Anderson:2009nt}. The space of homomorphisms from $\mathcal{F}$
to $V$, denoted $\op{Hom}_X(\mathcal{F},V)$, is then isomorphic to the space of global sections $H^0(X, \mathcal{F}^*\otimes V)$. If $V$ is an $SU(N)$ bundle, it is stable if all its sub-sheaves $\mathcal{F}$ have negative slope. Hence, we have that
\begin{equation}\label{eq:stability}
	\begin{gathered}
		V \text{ is stable}\\
		\Updownarrow\\
\mu(\mathcal{F})<0\quad\forall\,\mathcal{F} \text{ with } 
0<\rank \mathcal{F}<\rank V\text{ and } H^0(X, \mathcal{F}^*\otimes V)\neq0\ .
\end{gathered}
\end{equation}
Applying this statement to our case, the bundle $\mathcal{F}$ is a subbundle of $V_{\rep 3}^\prime$ if we can find a homomorphism 
$\mathcal{F}\hookrightarrow V^\prime_{\rep 3}$ or, equivalently, if
\begin{equation}
\label{eq:condition}
\op{Hom}_X(\mathcal{F},V^\prime_{\rep 3})=H^0(X,\mathcal{F}^*\otimes V^\prime_{\rep 3} )\neq 0\ .
\end{equation}
In the following, we will show that such a homomorphism does indeed exist. Let us start by tensoring with $\mathcal{F}^*$ the sequences
\begin{equation}
\label{eq:sequences_app3}
 0\rightarrow \mathcal{K}\rightarrow V^{\prime}_{\rep 3} \rightarrow W \rightarrow 0\ 
\end{equation}
and
\begin{equation}
\label{eq:sequences_app4}
 0\rightarrow \mathcal{F}\rightarrow W \rightarrow \mathcal{E} \rightarrow 0\ 
\end{equation}
to obtain
\begin{equation}
\begin{split}
\label{eq:sequences_app5}
& 0\rightarrow \mathcal{F}^*\otimes \mathcal{K}\rightarrow \mathcal{F}^*\otimes V^{\prime}_{\rep 3} \rightarrow \mathcal{F}^*\otimes W \rightarrow 0\ ,\\
& 0\rightarrow \mathcal{F}^*\otimes \mathcal{F}\rightarrow \mathcal{F}^*\otimes W \rightarrow \mathcal{F}^*\otimes \mathcal{E} \rightarrow 0\ .
\end{split}
\end{equation}
Taking long exact sequences in cohomology of these, gives
\begin{equation}
\begin{split}
\label{eq:12341}
 0&\rightarrow H^0(X,\mathcal{F}^*\otimes \mathcal{K})\rightarrow H^0(X,\mathcal{F}^*\otimes V^{\prime}_{\rep 3}) \rightarrow H^0(X,\mathcal{F}^*\otimes W )
 \\ 
 &\xrightarrow{\delta_1} H^1(X,\mathcal{F}^*\otimes \mathcal{K})\rightarrow \dots\ ,\\
 \end{split}
\end{equation}
and
\begin{equation}
\label{eq:12342}
\begin{split}
 0&\rightarrow H^0(X,\mathcal{F}^*\otimes \mathcal{F})\rightarrow H^0(X,\mathcal{F}^*\otimes W) \rightarrow H^0(X,\mathcal{F}^*\otimes \mathcal{E} )
 \\ 
 &\xrightarrow{\delta_2} H^1(X,\mathcal{F}^*\otimes \mathcal{F})\rightarrow \dots\ .\\
 \end{split}
\end{equation}

For a line bundle and its dual we have
\begin{equation}
\begin{split}
H^0(X,\mathcal{F}^*\otimes \mathcal{F})=H^0(X,\mathcal{O}_X)=\mathbb{C}\ ,\\
H^1(X,\mathcal{F}^*\otimes \mathcal{F})=H^1(X,\mathcal{O}_X)=0\ .
\end{split}
\end{equation}
Furthermore, if the line bundles $\mathcal{F}^*\otimes \mathcal{K}$ and $\mathcal{F}^*\otimes \mathcal{E}$ have negative
slopes somewhere in the Kähler cone, the zeroth cohomology classes
\begin{equation}
	H^0(X,\mathcal{F}^*\otimes \mathcal{K})=0\ ,\qquad H^0(X,\mathcal{F}^*\otimes \mathcal{E})=0\ ,
\end{equation}
vanish, as explained in Footnote 4 of \cite{Anderson:2012yf}. It can be shown that for our particular Schoen manifold, this condition is always satisfied if the line bundles $\mathcal{F}^*\otimes \mathcal{K}=L_1L_2^3$ and $\mathcal{F}^*\otimes \mathcal{E}=L_1^{-1}L_2^3$ have both positive and negative entries $m^i+3n^i$ and $-m^i+3n^i$, when written as $\mathcal{F}^*\otimes \mathcal{K}=\mathcal{O}_X(m^1+3n^1,m^2+3n^2,m^3+3n^3)$ and $\mathcal{F}^*\otimes \mathcal{E}=\mathcal{O}_X(-m^1+3n^1,-m^2+3n^2,-m^3+3n^3)$. This is generally the case for the line bundles we sample. In particular, it can be checked that the negative slopes condition is valid for the all the line bundle configurations from Table \ref{tab:solutions_branch1}.

Hence equations \eqref{eq:12341} and \eqref{eq:12342} become
\begin{equation}
\label{eq:delta_1_seq}
 0\rightarrow H^0(X,\mathcal{F}^*\otimes V^{\prime}_{\rep 3}) \rightarrow H^0(X,\mathcal{F}^*\otimes W )\xrightarrow{\delta_1} H^1(X,\mathcal{F}^*\otimes \mathcal{K})\rightarrow \dots\ ,\\
\end{equation}
and
\begin{equation}
\label{eq:C_seq}
 0\rightarrow \mathbb{C}\rightarrow H^0(X,\mathcal{F}^*\otimes W) \rightarrow 0
\end{equation}
respectively. 

From the sequence \eqref{eq:delta_1_seq} we learn that $H^0(X,\mathcal{F}^*\otimes V^{\prime}_{\rep 3})=\kernel \delta_1$. Therefore, to evaluate $H^0(X,\mathcal{F}^*\otimes V^{\prime}_{\rep 3})$, we must first analyze the coboundary map $\delta_1$. First note that from eq.~\eqref{eq:C_seq} we learn that $H^0(X,\mathcal{F}^*\otimes W)=\mathbb C$. Furthermore, we have that the fields $\tilde C_1$ are counted by $H^1(X,\mathcal{F}^*\otimes \mathcal{K})$. In the chosen vacuum branch all VEVs for the $\tilde C_1$ fields vanish. Since $\delta_1$ is determined by the vacuum state configuration, it follows that it maps only to the zero element of $H^1(X,\mathcal{F}^*\otimes \mathcal{K})$. Hence, $\kernel \delta_1=H^0(X,\mathcal{F}^*\otimes W)=\mathbb C$. 

 Putting this together, we conclude that
\begin{equation}
 H^0(X,\mathcal{F}^*\otimes  V^{\prime}_{\rep 3})=\mathbb C\ ,
\end{equation}
which is indeed non-zero. Therefore, according to \eqref{eq:condition}, there exists a homomorphism $\mathcal{F}\hookrightarrow V^{\prime}_{\rep 3}$ such that $\mathcal{F}$ is a subbundle of $V^{\prime}_{\rep 3}$.
Sequences \eqref{eq:sequences_app1} and \eqref{eq:sequences_app2} then tell us that both $\mathcal{F}$ and $\mathcal{K}$ are subbundles of $V_{\rep 3}^\prime$. Therefore, according to \eqref{eq:stability}, $V_{\rep 3}^\prime$ is stable only if
\begin{equation}
\mu({\mathcal{F}})<0 \quad\text{and}\quad \mu({\mathcal{K}})<0\ .
\end{equation}

\bibliographystyle{utphys}
\bibliography{citations2,extra}

\end{document}